\RequirePackage{fix-cm}
\documentclass[smallextended]{svjour3}      
\smartqed 
\usepackage{graphicx}
\usepackage{amsmath}
\usepackage{amssymb}
\usepackage[]{natbib}  
\usepackage{multirow}
\usepackage[utf8]{inputenc}  
\usepackage[T1]{fontenc}  
\usepackage{placeins}
\usepackage{hyperref}
\hypersetup{
    colorlinks=true,
    linkcolor=blue,
    filecolor=blue,      
    urlcolor=blue,
    citecolor=blue,
}

\journalname{Experimental Astronomy}

\begin{document}

\title{Ultra-Long Gamma-Ray Bursts detection with SVOM/ECLAIRs}

\titlerunning{ulGRBs with SVOM/ECLAIRs}  
\author{Nicolas Dagoneau \and
        Stéphane Schanne \and
        Jean-Luc Atteia \and
        Diego Götz \and
        Bertrand Cordier
}
\authorrunning{N. Dagoneau} 
\institute{N. Dagoneau, S. Schanne, D. Götz, B. Cordier \at CEA Paris-Saclay, IFRU/Département d'Astrophysique - AIM, 91191 Gif-sur-Yvette, France\\
\email{nicolas.dagoneau@cea.fr}
\and J-L. Atteia \at IRAP, Université de Toulouse, CNES, CNRS, UPS, Toulouse, France
}

\date{Received: 24 January 2020 / Accepted: 27 May 2020}
\maketitle

\begin{abstract}

Ultra-long Gamma-Ray Bursts (ulGRBs) are Gamma-Ray Bursts (GRBs) with an unusually long emission in X and gamma rays, reaching durations of thousands of seconds.
They could form a specific class of high-energy transient events, whose origin is still under discussion.
The current sample of known ulGRBs consists of a few tens of events which have been detected so far by the Burst Alert Telescope (BAT) aboard the Neil Gehrels Swift Observatory and some other instruments. 
The SVOM mission which is scheduled to begin operations after 2021 could help to detect and observe more ulGRBs thanks to its soft gamma-ray telescope ECLAIRs. 
After an introduction on ulGRBs and the SVOM mission, we present the results of our simulations on the capabilities of ECLAIRs to detect ulGRBs.
First we use the sample of ulGRBs detected by Swift/BAT and simulate these events through a model of the instrument and the prototype trigger software that will be implemented onboard ECLAIRs.
Then we present a study of the ECLAIRs capabilities to detect a synthetic population of ulGRBs built by transporting the ulGRBs detected by Swift/BAT to higher redshifts.
Finally we give an estimate of the ulGRB rate expected to be detected by ECLAIRs and show that ECLAIRs can detect at least as much ulGRBs as BAT.


\end{abstract}

% --------------------------------------------
\section{Introduction}
% --------------------------------------------

Gamma-ray bursts (GRBs) appear in the gamma-ray and hard X-ray photon energy range as bright transient point-like sources at random positions on the sky.
They are usually classified based on their duration in that energy range into two distinct families \citep{kouveliotou_identification_1993, dezalay_hardness-duration_1996}.
The so-called short bursts (sGRBs) with a duration of less than 2 s are believed to be the result of the coalescence of two compact objects such as neutron stars, whereas the long bursts (lGRBs) with a typical duration of about 30 s are produced during the collapse of peculiar kinds of very massive stars.
Recently, it has been pointed out that some lGRBs show an extraordinarily long duration of more than 1000 s.
One of these bursts is GRB 111209A \citep{gendre_ultra-long_2013} which was active for about 25000 s.
GRBs with such highly atypical durations have been called ultra-long gamma-ray bursts (ulGRBs) \citep{levan_new_2013}. 
They could form a new third class of GRBs.
One explanation which has been proposed for their ultra-long duration is that they could have progenitors different than classical GRBs: they could be produced either by the core collapse of a low-metallicity supergiant blue star \citep{gendre_ultra-long_2013}, the birth of a magnetar following the collapse of a massive star \citep{greiner_very_2015} or the collapse of a Pop III star \citep{nakauchi_long-duration_2012, kinugawa_long_2019}.
However ulGRBs could also just represent the tail of the standard long GRB distribution \citep{virgili_grb091024a_2013}.
In any case, it is clear that the durations of these bursts make them so peculiar that they deserve further studies.
Their fast detection and follow-up is essential to better understand them and to answer some questions, such as: do they really form a new class of GRBs, what are their progenitors, what is their central engine, what is the mechanism of such a long-time energy injection responsible of the production of their long-lasting gamma-ray and X-ray emission?
Additionally it is interesting to notice that the famous ulGRB, GRB 111209A, was associated with a very luminous supernova at redshift z=0.677 \citep{greiner_very_2015}.
This is exactly the kind of phenomenon that requires a multi-wavelength follow-up from the detection of the burst on, and over several days.
As of today, the few well studied ulGRBs have been detected by the Burst Alert Telescope (BAT) \citep{barthelmy_burst_2005} onboard the Neil Gehrels Swift Observatory. 

The future SVOM mission will be dedicated to gamma-ray bursts and transient astronomy, and may contribute to detect ulGRBs and to better understand them, especially with its image trigger. Also, thanks to its follow-up capabilities in the X-ray, optical and near-infra-red bands, SVOM could provide a good insight into the properties of this new class of events.

In Sec.~\ref{sec:svom} we describe the SVOM mission and its hard X-ray coded-mask telescope ECLAIRs, equipped with its onboard GRB-trigger system.
We present in Sec.~\ref{sec:sample} the sample of ulGRBs used to study their detection by the ECLAIRs telescope.
In Sec.~\ref{sec:simulation} we present our simulations of those ulGRBs projected through a model of ECLAIRs as well as the performances of the prototype GRB-trigger on this sample in terms of detection timescale, localisation accuracy and most efficient energy-band used for triggering. In Sec.~\ref{sec:rates} we present the ulGRB rate expected for ECLAIRs.

% --------------------------------------------
\section{The SVOM mission}
\label{sec:svom}
% --------------------------------------------

\begin{figure}[h!]
\centering
\includegraphics[width=\textwidth]{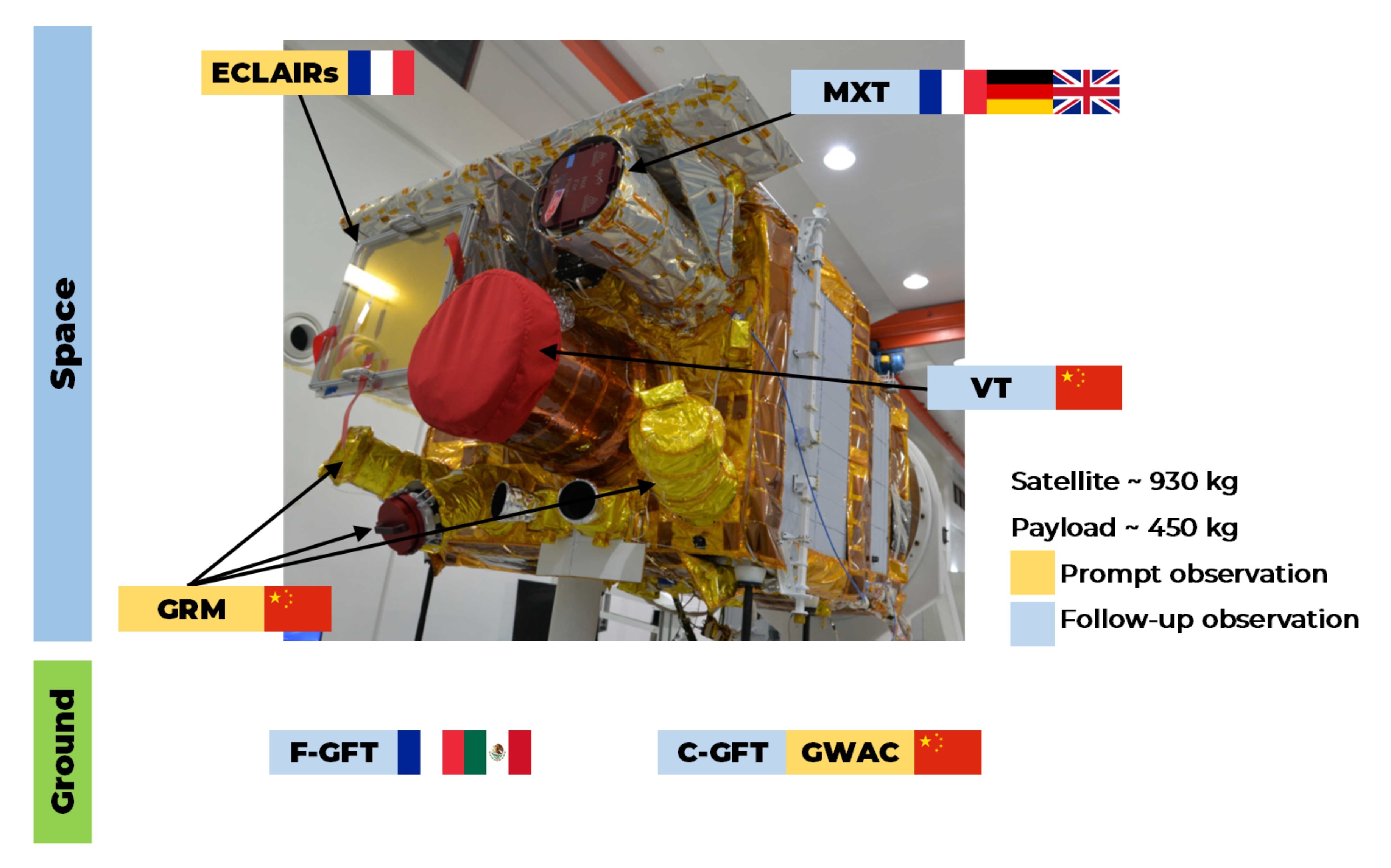}
\caption{Overview of the SVOM platform. The wide field of view instruments are shown in orange boxes. The small field of view instruments are depicted in blue boxes. The photo shows the structural and thermal models of the instruments on the qualification model of the satellite (CNES/SECM).}
\label{fig:svom}
\end{figure}

The SVOM (Space-based multi-band astronomical Variable Objects Monitor, \citealt{wei_deep_2016}) is a French-Chinese mission dedicated to the detection and follow-up of GRBs and high-energy transients, currently under development and planned to be operational after 2021.
Figure \ref{fig:svom} shows the SVOM platform including four space instruments: the Gamma Ray-burst Monitor (GRM), the ECLAIRs telescope, the Microchannel X-ray Telescope (MXT) and the Visible Telescope (VT), and the additional dedicated ground instruments: the Ground Wide Angle Cameras (GWAC) and the two Ground Follow-up Telescopes (GFTs).

The goal of the SVOM core-program is the multi-wavelength study of GRBs, including the prompt and the afterglow emission, in order to obtain a complete sample of well-characterised GRBs.  
The GRB prompt emission is first detected and its gamma-ray spectrum is characterised with the wide-field-instruments ECLAIRs and GRM onboard, and simultaneously studied in the visible band with the GWACs on-ground, which points as much as possible to the same sky region.
After the GRB detection and localisation onboard by ECLAIRs, the spacecraft will slew to the GRB position, and the afterglow will be observed by the narrow-field-instruments, MXT in X-rays and VT in the visible band. 
Simultaneously the GRB alert including its position will be transmitted to the ground through a VHF antenna, and the GFTs will perform follow-up observations in the visible and near-infrared bands.
The GRB alerts and their subsequent refined positions obtained by the SVOM follow-up telescopes are quickly and publicly distributed to the whole community of interested observers of the transient sky.
This strategy will enhance the number of accurately localised bursts, which is essential to perform fruitful follow-up observations with large ground-based facilities, including spectroscopic telescopes, important to determine the redshift of those events.
With the SVOM strategy to perform observations mainly towards the night sky on Earth, most of the SVOM-detected burst will be observable immediately after detection with large spectroscopic ground-based telescopes located in the Earth tropical zones.

In this paper we will focus only on the onboard detection of ulGRBs in the soft $\gamma$-ray band by ECLAIRs, whereas the questions related to their follow-up observations or to the ground off-line detection is left aside for further studies.

\begin{figure}[h!]
\centering
\includegraphics[width=\textwidth]{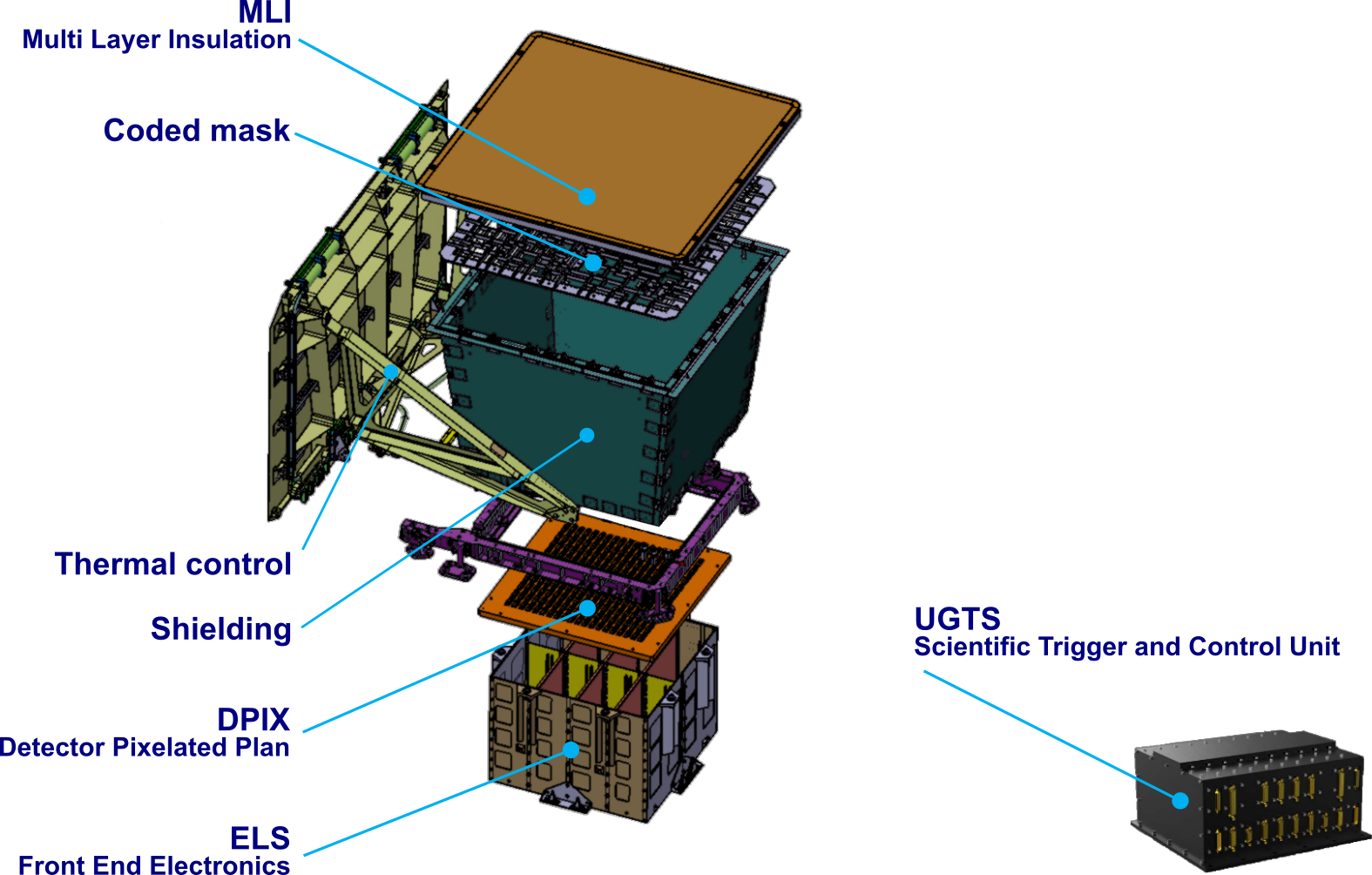}
\caption{Schematic representation of the ECLAIRs telescope onboard SVOM. }
\label{fig:eclairs}
\end{figure}

The ECLAIRs space telescope onboard SVOM, dedicated to GRB detection, is a coded-mask aperture telescope operating in the energy range from $4$ to $150$ keV \citep{takahashi_x-gamma-ray_2014}.
With such a low energy threshold, it is particularly well suited for the detection of X-ray rich GRBs and highly redshifted GRBs.
Figure \ref{fig:eclairs} shows a schematic representation of ECLAIRs.
Its detection plane DPIX is composed of 80$\times$80 CdTe pixels of thickness 1 mm and 4$\times$4 mm$^2$ active surface each, arranged on a 4.5 mm grid.
Its self-supporting tantalum mask has dimensions 54$\times$54 cm$^2$ and 0.6 mm thickness, providing imaging capabilities up to 120 keV.
With its mask-detector distance of 46 cm, the total field of view is 2 sr and the localisation accuracy of sources at detection limit amounts to about 12 arcmin.

As the ECLAIRs tantalum mask is opaque to soft $\gamma$-ray photons, a source emitting such photons will project the shadow of the mask pattern onto the detector, and the image recorded by the detector is called a shadowgram.
The sky image is built from the shadowgram using the mask deconvolution method \citep{goldwurm_integral/ibis_2003}.

The ECLAIRs telescope is also affected by different background components \citep{zhao_influence_2012,mate_simulations_2019}.
In this study we consider only the dominant Cosmic X-ray Background (CXB) described by \cite{moretti_new_2009}.
In ECLAIRs, due to the geometry and the large field of view of the instrument, the CXB produces a non-flat close-to-quadratic shape on the detection plane that needs to be removed prior to deconvolution. 
Moreover, some known X-ray sources (Sco X-1, Crab, Cyg X-1, ...) can be present in the field of view. 
In sum, this leads to the superposition of different shadows in the detector-shadowgram and to coding noise after deconvolution.
In order to enhance the GRB detection capabilities, strong known-sources also have to be cleaned.
The detailed cleaning methods used and their performances as well as the known-source catalogue developed for ECLAIRs will be presented in other papers in preparation. 
In this paper, we assume in our simulations that neither the Earth nor sources are present in the field of view, and we consider only the presence of a GRB and the CXB background.
Also in this paper we clean the CXB background by fitting and subtracting a quadratic-shaped model of the CXB from the shadowgram prior to deconvolution.

The ECLAIRs telescope uses two different simultaneously-running trigger algorithms: a count-rate trigger monitoring significant rate increases on timescales from $10$ ms up to $20.48$ s followed by the imaging of the corresponding timescale in excess, and an image trigger which builds $20.48$ seconds-long sky images, which are stacked to form and analyse systematically all sky images on timescales from $20.48$ s up to $\sim 20$ min \citep{schanne_scientific_2013, schanne_eclairs_2015}.
We use the prototype trigger software in order to study its detection capabilities of the ulGRBs.

\begin{figure}[h!]
\centering
\includegraphics[width=.9\textwidth]{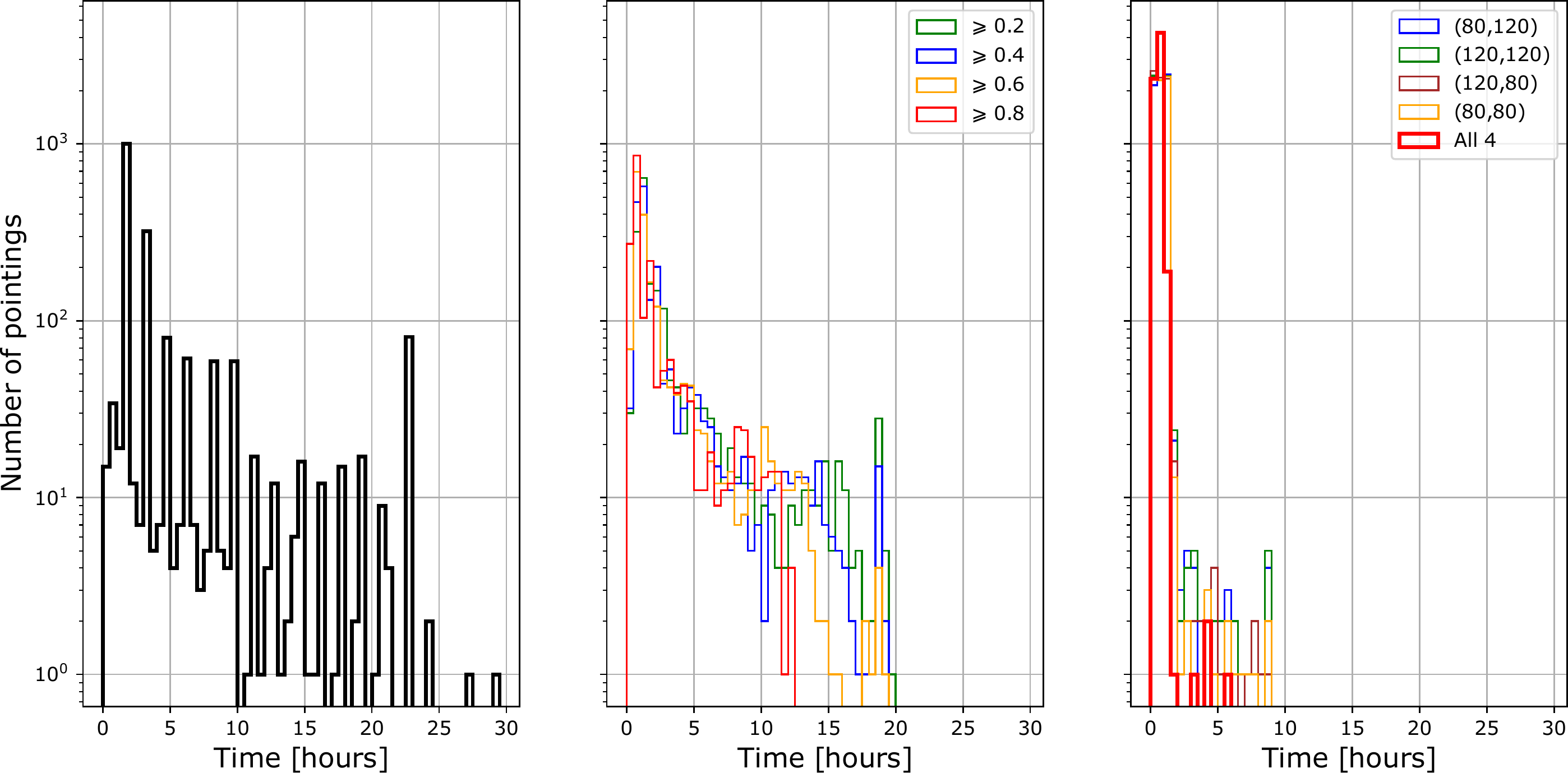}
\caption{Distribution of the SVOM pointing and observation durations over one year (CNES simulation). \textbf{Left}: stable pointings (possibly interrupted by SAA crossings and Earth occultations). The high peaks represent durations which are multiples of a full orbit (about 90 min duration). 
Many stable pointings of up to 14 orbits (about 22 h) are foreseen. \textbf{Centre}: interrupted observations (pointing durations with subtracted interruption durations due to SAA crossings and Earth occultations). The different coloured histograms represent different ratios of the field of view free of Earth. \textbf{Right}: continuous observations between slews, SAA crossings and Earth occultations for each of the four corners of the fully coded field of view, as well as for all four corners together. The median of the red histogram (fully coded field of view free of Earth) is 43 min. Few long observations (up to 5 hours) in which the fully coded field of view remains free of Earth are possible.}
\label{fig:pointings}
\end{figure}

The SVOM spacecraft follows roughly an anti-solar pointing law in order to ensure that its field of view is simultaneously observable by ground instruments in the night hemisphere of the Earth.
This leads to long stable pointings (up to 10 hours and more, as shown in Fig. \ref{fig:pointings}) at the expense of Earth passages through the field of view and variable background along the orbit resulting mainly from the CXB modulated by the Earth passage.
ECLAIRs benefits from this pointing-law, which allows it to perform long-exposure imaging with its onboard image-trigger, up to 20 minutes, the longest timescale currently foreseen.

\begin{figure}[h!]
\centering
\includegraphics[width=.9\textwidth]{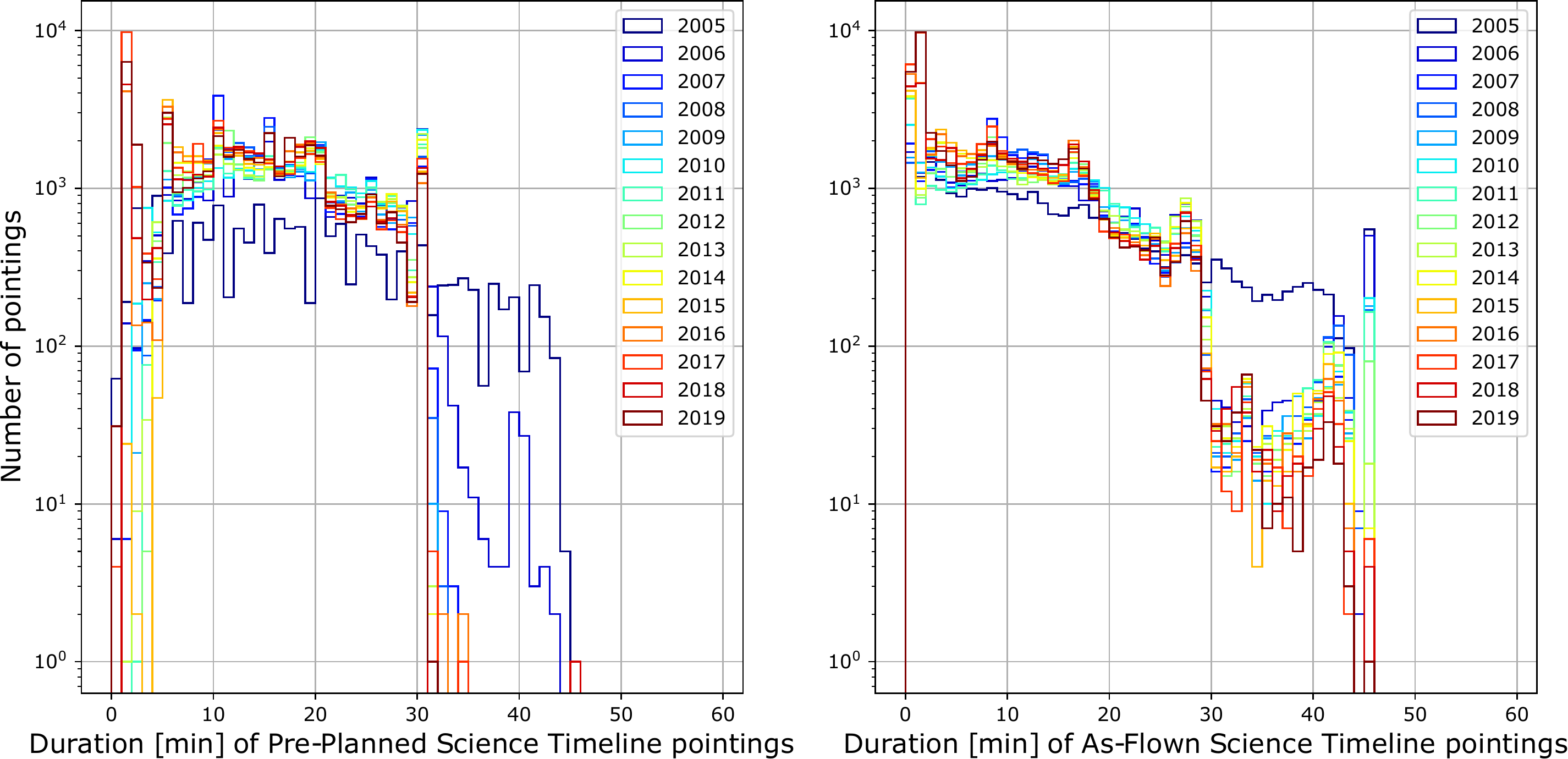}
\caption{Distribution of the Swift pointing durations over each year of observation between 2005 and 2019 (data from \url{https://www.swift.psu.edu/operations/obsSchedule}). \textbf{Left}: pointings from the Pre-Planed Science Timeline which does not include GRB follow-up and ToOs. \textbf{Right}: pointings from the As-Flown Science Timeline which includes GRB follow-up and ToOs, leading to pointings between 30 min and 45 min.}
\label{fig:swift_pointings}
\end{figure}

For comparison, Fig. \ref{fig:swift_pointings} shows the distribution of the pointing durations for Swift, with the longest stable pointing being $\sim$ 30 min, which ensures to thermally stabilise the XRT, and also to keep the Earth outside the field of view of BAT as much as possible.
Therefore however, any exposure on a weak new source will be interrupted by a slew (possibly before its detection) as soon as the Earth enters the edge of the field of view, and most likely before the Earth actually occults the source.
On the contrary, SVOM, with its long pointing durations, will let the Earth enter deeply into the ECLAIRs field of view, and a possible new source will only become unobservable when it actually will be occulted by the Earth.
Furthermore the presence of the Earth in the field of view will also reduce part of the sky background (CXB) projected onto the detector, enhancing its capability to find faint sources on the part of the field of view non occulted by the Earth.
The ECLAIRs onboard triggering system is designed to work under the variable background condition when the Earth is partially present in the field of view.

The detection of transients on timescales longer than 20 min, which can be performed on the ground, are not discussed here.

% --------------------------------------------
\section{The ultra-long Gamma-Ray Bursts sample}
\label{sec:sample}
% --------------------------------------------

We built a sample of ultra-long GRBs in order to study their detection with the ECLAIRs prototype trigger software.
The sample is built using the Swift BAT GRB catalogue summary tables\footnote{\url{https://swift.gsfc.nasa.gov/results/batgrbcat/summary_cflux/summary_GRBlist/list_ultra_long_GRB_comment.txt}} and some complements from the literature.
Bursts are reported in Table~\ref{table:ulGRBs}.

\begin{figure}[ht!]
\centering
\includegraphics[width=\textwidth]{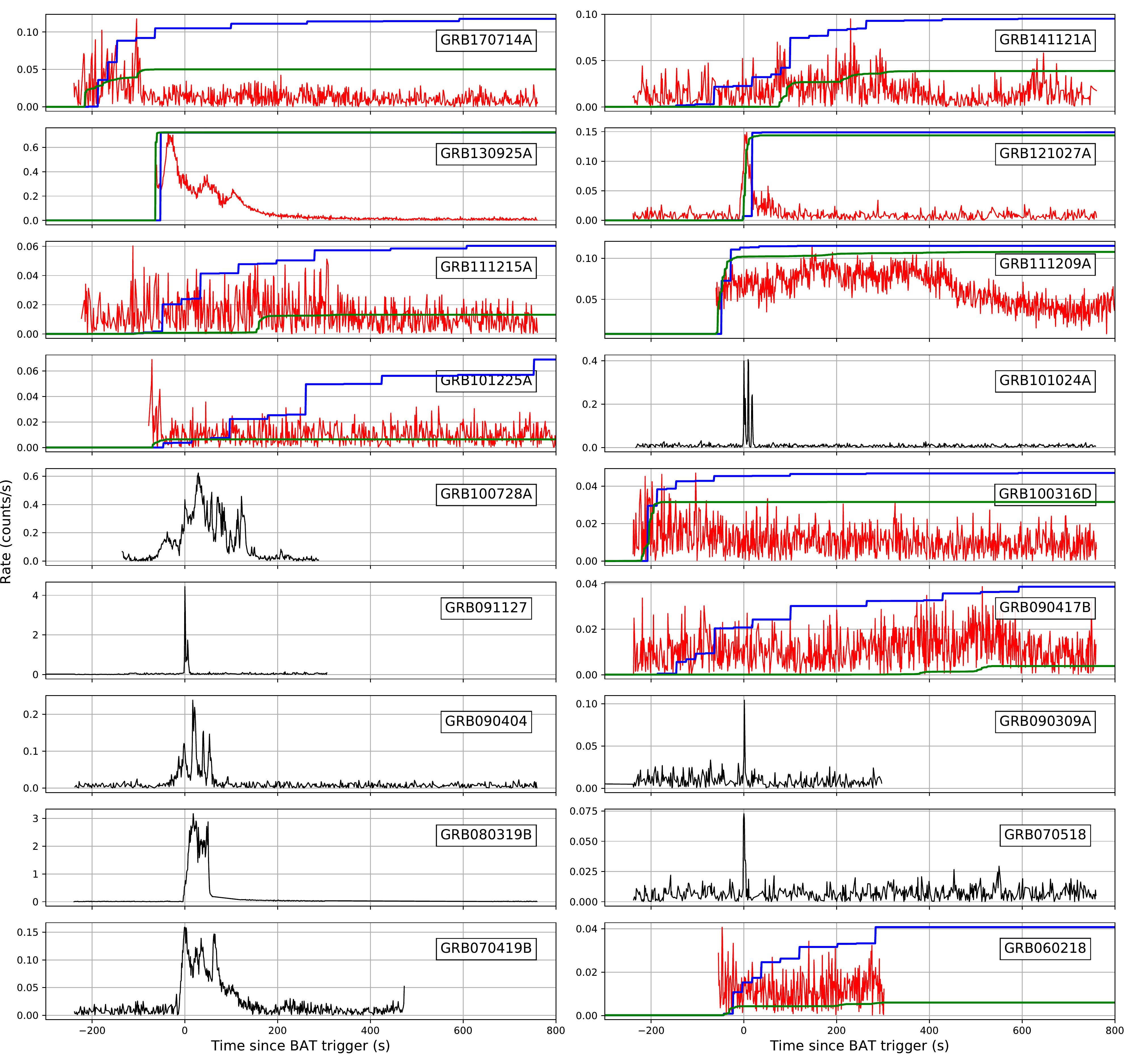}
\caption{Lightcurves of the ultra-long GRB sample from Swift BAT in 15-150 keV in counts/s. Red curves correspond to the 10 ulGRBs of our sample. Blue and green curves show the cumulative distributions of the first detection time over the 1000 simulations for the image trigger (blue) and the count-rate trigger (green) in the total field of view.}
\label{fig:lc_ulGRB}
\end{figure}

The sample is composed of 19 ulGRBs with BAT data.
However, the bursts with BAT lightcurves given in Fig. \ref{fig:lc_ulGRB} have to be examined one by one to confirm their belonging to that sample.
Indeed, some of them do not look especially long or are not mentioned in the literature as ulGRB (except in the Swift summary table). 
We will keep the bursts in the sample (in bold in the table) under the condition that the lightcurve shows ultra-long emission or the burst is mentioned as a ulGRB in the literature. 

\begin{table}[h]
\begin{center}
\begin{tabular}{ccccccc}
\hline
Name & $T_{90}$\textsuperscript{a} & $T_{90\text {,err}}$\textsuperscript{a} & refined $T_{90}$ & $T_X$\textsuperscript{b} & $T_{\text{burst}}$\textsuperscript{c} & z\textsuperscript{a}  \\ \hline\hline
\textbf{170714A} & \textbf{459} & \textbf{94} &  & \textbf{16600} &  & \textbf{0.793} \\ 
\textbf{141121A} & \textbf{481} & \textbf{38} &  & \textbf{< 5000} &  & \textbf{1.47} \\ 
\textbf{130925A} & \textbf{160} & \textbf{3} & \textbf{5000\textsuperscript{d}} & \textbf{10000} & \textbf{11641} & \textbf{0.348} \\
\textbf{121027A} & \textbf{80} & \textbf{40} & \textbf{6000\textsuperscript{d}} & \textbf{8000} & \textbf{35399} & \textbf{1.773} \\
\textbf{111215A} & \textbf{373} & \textbf{93} &  & \textbf{990} & \textbf{1462} &  \textbf{2.1\textsuperscript{f}} \\
\textbf{111209A} & \textbf{810} & \textbf{52} & \textbf{13000\textsuperscript{d}} & \textbf{25400} & \textbf{63241} & \textbf{0.677} \\ 
\textbf{101225A} & \textbf{> 1377} &  & \textbf{7000\textsuperscript{d}} & \textbf{5296} & \textbf{106659} & \textbf{0.847} \\ 
101024A & 18 & 0 &  &  &  &  \\ 
100728A & 193 & 10 &  &  & 931 & 1.567 \\ 
\textbf{100316D} & \textbf{521} & \textbf{439} & \textbf{1300\textsuperscript{e}} &  & \textbf{1300} & \textbf{0.0591} \\ 
091127 & 6 & 0 &  &  & 5559 & 0.49044 \\ 
091024 & 112 & 13 & 1300\textsuperscript{e} &  &  & 1.092 \\ 
\textbf{090417B} & \textbf{266} & \textbf{35} & \textbf{2130\textsuperscript{e}} & \textbf{535} & \textbf{2098} & \textbf{0.345} \\ 
090404 & 82 & 14 &  &  & 242 & 2.87 \\ 
090309A & 3 & 1 &  &  &  &  \\ 
080407 &  &  & 2100\textsuperscript{e} &  &  &  \\ 
080319B & 124 & 3 &  &  &  & 0.9382 \\ 
070518 & 5 & 0 &  &  & 357 & 1.161(?) \\ 
070419B & 238 & 14 &  &  & 387 & 1.9588 \\ 
060814B &  &  & 2700\textsuperscript{e} &  &  &  \\ 
\textbf{060218} & & & \textbf{2100\textsuperscript{e}} & \textbf{2917} & \textbf{11830} & \textbf{0.03342} \\ \hline
\end{tabular}
\caption{Ultra-long GRB sample (durations in s). GRBs written in bold are those with BAT data used for our simulations. The redshift of GRB 070518 (1.161) reported in the Swift table from D. Perley's Keck GRB Host project conflicts with z$<$0.7 limit from GCN 6419. \label{table:ulGRBs}}
\end{center}
\textsuperscript{a} From Swift BAT GRB catalogue summary tables (\url{https://swift.gsfc.nasa.gov/results/batgrbcat/index_tables.html}); bursts with no data have been detected by other missions\\
\textsuperscript{b} From \cite{gendre_can_2019}: value defined by \cite{boer_are_2015} and based on Swift XRT ligthcurves\\
\textsuperscript{c} From \cite{zhang_how_2014}: based on Swift XRT ligthcurves\\
\textsuperscript{d} Refined $T_{90}$ from \cite{levan_swift_2015}\\
\textsuperscript{e} Approximate duration from
\cite{virgili_grb091024a_2013} (Table 6)\\
\textsuperscript{f} From \cite{van_der_horst_detailed_2014}\\
\end{table}

\begin{itemize}
\item GRB 141121A is missing in the Swift summary table. However we included it in our sample since it is described as ulGRB in \cite{cucchiara_happy_2015} and its BAT lightcurve shows ultra-long emission.

\item GRB 130925A has incomplete BAT lightcurve due to spacecraft manoeuvres \citep{evans_grb130925a:_2014}, however we chose to keep it in the sample since it reflects real emission of an ulGRB and it belongs to the ``gold'' sample of \cite{gendre_can_2019}.

\item GRB 121027A is classified as ulGRB because of its long lasting X-ray emission \citep{zhang_how_2014}. Its $\gamma$-ray lightcurve is more typical of classical long-GRBs. We keep it in the sample for the same reasons as previously.

\item GRB 101024A and GRB 100316D have been detected in BAT survey data before the BAT trigger \citep{lien_third_2016}. We only keep GRB 100316D since its lightcurve is similar to the one of GRB 101225A and it is reported as an ulGRB in the literature \citep{gendre_ultra-long_2013}.

\item GRB 100728A, GRB 090404, GRB 090309A, GRB 070518 and GRB 070419B are only reported in the Swift summary table and their lightcurves do not show ultra-long emission, hence they are removed from the sample.

\item GRB 091127 is also removed from the sample. Despite the high $T_{\mathrm{burst}}$ value due to the association with SN2009nz \citep{cobb_discovery_2010}, it does not show ultra-long structures in the BAT lightcurve and is not reported as ulGRB in the literature (except in the Swift summary table).

\item GRB 091024 is removed (no data until $\sim T_0 + 3000$ s possibly due to Earth constraint).

\item GRB 080319B is the ``naked-eye'' GRB \citep{racusin_grb080319b:_2008}. This may be the reason why it is reported as a ulGRB in the Swift summary table due its brightness. It is removed from sample.

\item GRB 060218 has been detected by Swift BAT and belongs to the sample but $T_{90}$ is not known because it is longer than the duration of event data. It belongs to the ``silver'' sample of \cite{gendre_can_2019}.
\end{itemize}

\FloatBarrier

About the burst duration, we can notice the existence of various definitions.
The intrinsic duration in the observer frame is difficult to evaluate for such events since its measurement depends on the instrument sensitivity and its capability to monitor the gamma-ray emission for a long period of time but also on the brightness of the burst.

Since Swift repoints frequently (several times per orbit), a stable pointing cannot be ensured for a long time, causing BAT to miss part of the ulGRB emission.
Furthermore, the Swift XRT has shown that the central engine is active for a longer time than $T_{90}$ measured by BAT \citep{levan_swift_2015}, thanks to the much higher sensitivity of XRT compared to BAT (in soft X-rays).
Therefore the BAT $T_{90}$ does not reach $>$ 1000 s for most ulGRBs.

\cite{zhang_how_2014} and \cite{boer_are_2015} propose different definitions of the activity time (respectively $T_\mathrm{burst}$ and $T_\mathrm{x}$) based on the X-ray emission. In fact, the so-called ``ultra-long'' bursts studied in the literature do not systematically have ultra-long durations in the gamma-ray domain, but have very long activity durations, since the X-ray flashes and internal plateaus are interpreted as manifestations of the central engine activity.
The refined $T_{90}$ from \cite{levan_swift_2015} differs from the standard analysis $T_{90}$ to take into account the observations during multiple orbits.
The one from \cite{virgili_grb091024a_2013} is an approximation of the gamma emission including quiescence or low-level emission.
More recently, \cite{gendre_can_2019} shows that there is a ``grey zone'' between $1000 < T_X < 5000$ where it is difficult to distinguish between classical long bursts and ulGRBs, and the probability for a burst to be ultra-long with $T_X$ = 1000 is 0.4167.

In the end, we kept 10 ulGRBs in our sample (in bold in the table).
This sample will be used for simulations through the ECLAIRs imaging and trigger algorithms presented in Sec.~\ref{sec:simulation}.
Despite the multiple definitions of the duration, they all reach 1000 s in one of the values (except for GRB 170714A and GRB 141121A which have been detected more recently).
We can also notice that the redshifts are not especially high, which makes it difficult to link them with the Pop III stars \citep{kinugawa_long_2019}.
However, knowing the redshift, we are able to transport lightcurves and spectra to higher redshifts and to study the detection capabilities of possible yet undetected high-z ulGRBs by ECLAIRs.

\begin{figure}[h!]
\centering
\includegraphics[width=.8\textwidth]{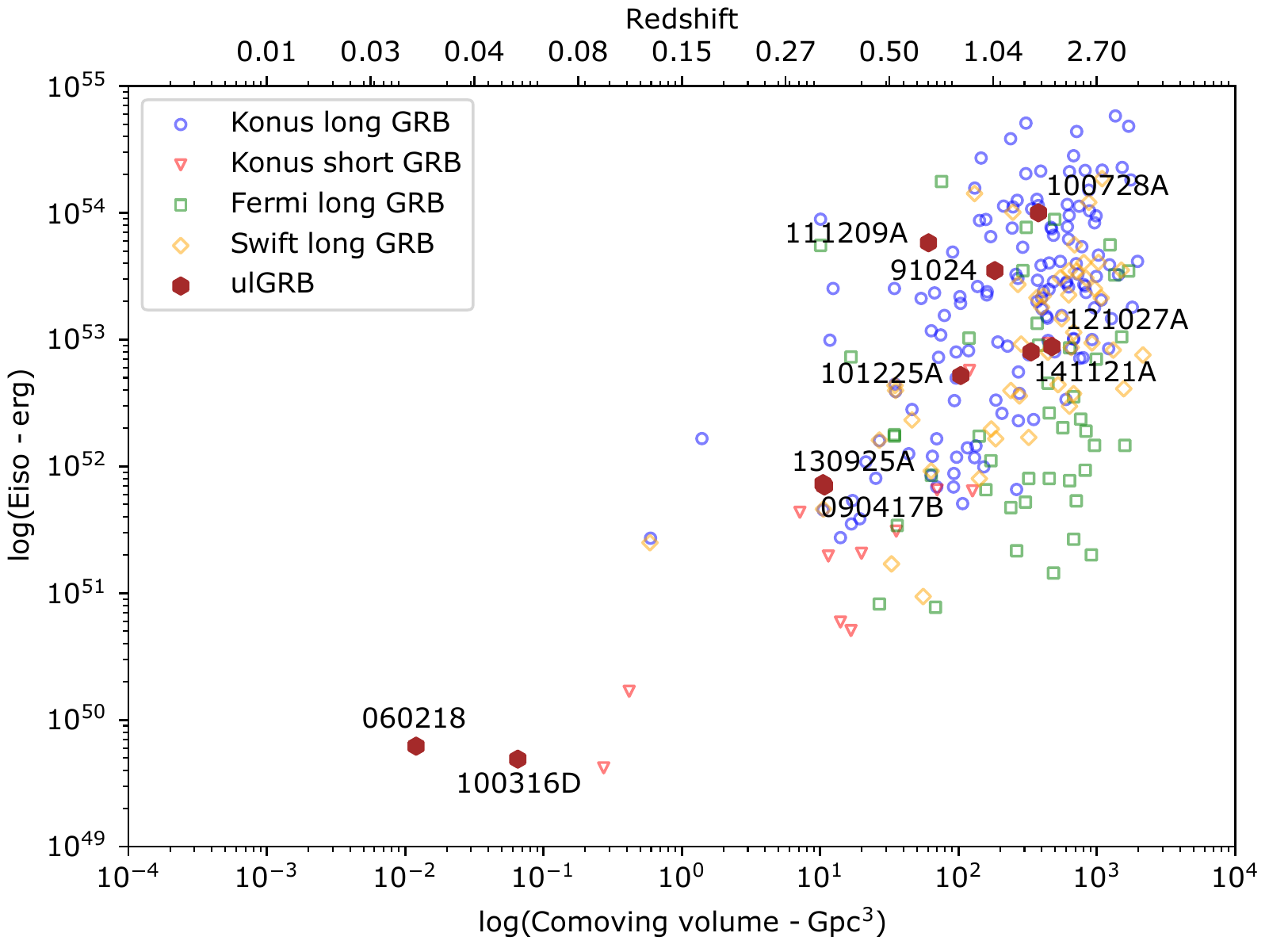}
\caption{Distribution of GRBs in the plane log(Comoving volume) - log($E_{\mathrm{iso}}$). Konus bursts are taken from \cite{tsvetkova_konus-wind_2017}, Fermi long bursts from \cite{heussaff_epeak_2013} and Swift long bursts from \cite{nava_complete_2012}.}
\label{fig:Eiso_V}
\end{figure}

Figure \ref{fig:Eiso_V} gives the distribution of GRBs in the plane $E_{\mathrm{iso}}$ versus comoving volume for short and long bursts, additionally to our ulGRBs with known redshift and isotropic energy.

We remark that GRB 060218 and GRB 100316D are quite different from the other ulGRBs. They are probably connected to supernova shock-breakout events \citep{campana_association_2006,waxman_grb_2007,starling_discovery_2011}. GRB 090417 and GRB 130925A are very similar despite a large separation on the sky of about $\sim 157 \deg$. Both emissions are interpreted as being scattered by dust in the line of sight by \cite{holland_grb090417b_2010} and \cite{evans_grb130925a:_2014}.

% --------------------------------------------
\section{Simulation through ECLAIRs}
\label{sec:simulation}
% --------------------------------------------

In order to study the detection of the 10 ulGRBs by SVOM/ECLAIRs, we use our ray-tracing simulation software and the trigger prototype software \citep{schanne_scientific_2013}.
The ulGRB data were retrieved from the Swift BAT archive. Each event has been reprocessed by the task \textit{batbinevt} (v1.48) to create the lightcurves and the spectra.
We used the version 6.22 of the FTOOLS with the last version of the Swift calibration database (BAT 20171016).
The task \textit{batdrmgen} (v3.6) is used to compute the response matrix.
Spectra are fitted with a power law using XSPEC 12 over a maximum duration of 1300 seconds.
Spectra obtained from BAT data are defined in 15-150 keV and extrapolated down to 4 keV for simulation through ECLAIRs.
The spectrum photon indexes and fluxes in 4-150 keV are given in Table~\ref{table:powerlaw}, they may differ from \cite{gendre_can_2019} who computed them only for the first 300 seconds.
Large errors on the fluxes are due to the spectrum fit over a long time.
For the propagation of the ulGRBs photons through ECLAIRs we will use the mean fitted flux. 

\begin{table}[h]
\begin{center}
\begin{tabular}{c|c|c|c|c}
\hline
Name & Photon index & Flux (ph/cm$^2$/s) & Reduced $\chi^2$ & Duration (s) \\ \hline\hline
170714A & -1.71 $\pm$ 0.15 & 0.15 $\pm$ 0.10 & 1.18 & 1202 \\ 
141121A & -1.71 $\pm$ 0.11 & 0.20 $\pm$ 0.09 & 1.16 & 6097 \\ 
130925A & -1.89 $\pm$ 0.01 & 1.81 $\pm$ 0.10 & 1.55 & 16613 \\ 
121027A & -1.86 $\pm$ 0.16 & 0.09 $\pm$ 0.07 & 0.90 & 1202 \\ 
111215A & -1.70 $\pm$ 0.12 & 0.16 $\pm$ 0.08 & 1.49 & 13727 \\ 
111209A & -1.48 $\pm$ 0.01 & 1.12 $\pm$ 0.05 & 0.92 & 1023 \\ 
101225A & -2.00 $\pm$ 0.22 & 0.11 $\pm$ 0.10 & 1.03 & 10943 \\ 
100316D & -2.30 $\pm$ 0.09 & 0.33 $\pm$ 0.12 & 0.92 & 1202 \\ 
090417B & -1.81 $\pm$ 0.07 & 0.20 $\pm$ 0.06 & 0.88 & 17235 \\ 
060218 & -2.18 $\pm$ 0.12 & 0.26 $\pm$ 0.12 & 1.25 & 602 \\ \hline
\end{tabular}
\caption{Spectral properties of the 10 ulGRBs of our sample. Fluxes are given in the ECLAIRs band 4-150 keV. \label{table:powerlaw}}
\end{center}
\end{table}

After data pre-processing, we draw photons for each ulGRB during a maximum duration of 1300 s, with the arrival time according to the lightcurve and the energy according to the spectrum.
These photons are ray-traced through the ECLAIRs coded mask, starting from a position chosen randomly from an isotropic distribution in the total field of view considered by the prototype trigger (1.82 sr).
The total effective field of view considered here is smaller than the geometrical one of ECLAIRs (2.05 sr) because of computational issues near the border of the field of view, that forces us to discard 10 pixels at its edges.
This issue will be solved in the final version of the trigger software, and will give access to the complete field of view.
The results are also given for the centre of the field of view (the fully coded field of view, where the mask shadow fully intercepts the detector, 0.15 sr) in which the sensitivity is the best.
Each of the ulGRBs is simulated at 1000 different positions in the total field of view.
Additionally, Cosmic X-ray Background (CXB) photons \citep{moretti_new_2009} are also propagated through ECLAIRs and an internal background of 0.003 counts/s/cm$^2$/keV is added \citep{sizun_synthesis_2011}.
We use a batch of 100 simulations of the CXB for each of the ulGRBs.
The $i$th simulation of each ulGRB uses a random simulation of the CXB among the 100 possibilities.
Finally, the counts recorded by the detector (after the ray-tracing through the mask and the energy redistribution with detector inefficiencies taken into account) are processed by the trigger prototype software.
The trigger returns for each detected event information such as the timescale, the energy strip, the signal to noise ratio and the position on the sky.
Before further analysis presented in the next sections, we discard the events for which the euclidean distance between the injected position and the position computed by the trigger software is larger than 2 pixels in the sky (the sky pixel size ranges from 27 arcmin at the edge to 33 arcmin in the centre of the field of view). Those wrongly associated positions (less than 0.5$\%$ for all bursts of our sample) are found very close to the edge of our effective field of view.

% --------------------------------------------
\subsection{Detection timescale}
\label{subsec:timeslice}
% --------------------------------------------

We are interested in the first timescale needed to detect those bursts above the imaging threshold $\mathrm{SNR}=6.5$.
We index the timescales $n_{\mathrm{img}}=1..7$ for durations $2^{n_{\mathrm{img}}-1} \times 20.48$ s in the image trigger and $n_{\mathrm{cr}}=1..12$ for durations $2^{n_{\mathrm{cr}}-1} \times 10$ ms in the count-rate trigger). 

First, we will focus on the fully coded field of view.
Tab.~\ref{table:firts-trig-slice-tc-img} and Tab.~\ref{table:firts-trig-slice-tc-cnt} show the minimum duration needed to detect the ulGRBs with the image and count-rate trigger prototype softwares respectively.
The first result is that all of the 10 ulGRB are always detected by the image trigger, for all the simulated positions in the fully coded field of view. 

With the image trigger, GRB 130925A and GRB 111209A are always detected on the shortest timescale of 20.48 s. 
GRB 170714A and GRB 060218 are mostly detected on a timescale of 40.96 s.
GRB 090417B is mostly detected on a timescale of 81.92 s.
GRB 141121A and GRB 111215A are mostly detected on a timescale of 163.84 s.
The timescales of 327.68 s is mainly not used and the longest timescales of 655.36 s and 1310.72 s are not used at all.

\begin{table}[h]
\begin{center}
\begin{tabular}{c|c|c|c|c|c|c|c|c}
\hline
\multirow{3}{*}{Name} & \multicolumn{8}{c}{Index $n_{\mathrm{img}}$ and duration of the timescale (seconds)} \\ \cline{2-9}
& \multirow{2}{*}{no detect.} & 1 & 2 & 3 & 4 & 5 & 6 & 7 \\ \cline{3-9}
&            & 20.48 & 40.96 & 81.92 & 163.84 & 327.68 & 655.36 & 1310.72 \\ \hline
GRB 170714A & 0 & 9.5 & 83.2 & 7.4 & 0 & 0 & 0 & 0 \\
GRB 141121A & 0 & 1.3 & 5.2 & 13 & 80.5 & 0 & 0 & 0 \\
GRB 130925A & 0 & 100 & 0 & 0 & 0 & 0 & 0 & 0 \\
GRB 121027A & 0 & 97.1 & 1.5 & 0 & 1.5 & 0 & 0 & 0 \\
GRB 111215A & 0 & 2.5 & 7.5 & 16.2 & 72.5 & 1.2 & 0 & 0 \\
GRB 111209A & 0 & 100 & 0 & 0 & 0 & 0 & 0 & 0 \\
GRB 101225A & 0 & 1.1 & 23 & 19.5 & 49.4 & 6.9 & 0 & 0 \\
GRB 100316D & 0 & 79.8 & 20.2 & 0 & 0 & 0 & 0 & 0 \\
GRB 090417B & 0 & 0 & 19.1 & 65.2 & 15.7 & 0 & 0 & 0 \\
GRB 060218 & 0 & 11.9 & 83.3 & 4.8 & 0 & 0 & 0 & 0 \\ \hline
\end{tabular}
\caption{First detection timescale of the ulGRBs by the image trigger prototype software within the fully coded field of view (percentage of simulated bursts). \label{table:firts-trig-slice-tc-img}}
\end{center}
\end{table}

With the count-rate trigger, the for our burst sample detection occurs always on a timescale longer or equal to 0.64 s. 

\begin{table}[h]
\begin{center}
\begin{tabular}{c|c|c|c|c|c|c|c|c|c|c|c|c|c}
\hline
\multirow{3}{*}{Name} & \multicolumn{13}{c}{Index $n_{\mathrm{cr}}$ and duration of the timescale (milliseconds)} \\ \cline{2-14}
 & \multirow{2}{*}{no detect.} & 1 & 2 & 3 & 4 & 5 & 6 & 7 & 8 & 9 & 10 & 11 & 12 \\ \cline{3-14}
 &            & 10 & 20 & 40 & 80 & 160 & 320 & 640 & 1280 & 2560 & 5120 & 10240 & 20480 \\ \hline
GRB 170714A & 0 & 0 & 0 & 0 & 0 & 0 & 0 & 15.8 & 15.8 & 65.3 & 1.1 & 0 & 2.1 \\
GRB 141121A & 1.3 & 0 & 0 & 0 & 0 & 0 & 0 & 0 & 5.2 & 5.2 & 40.3 & 20.8 & 27.3 \\
GRB 130925A & 0 & 0 & 0 & 0 & 0 & 0 & 0 & 100 & 0 & 0 & 0 & 0 & 0 \\
GRB 121027A & 0 & 0 & 0 & 0 & 0 & 0 & 0 & 0 & 1.5 & 33.8 & 61.8 & 2.9 & 0 \\
GRB 111215A & 16.2 & 0 & 0 & 0 & 0 & 0 & 0 & 0 & 2.5 & 2.5 & 35 & 31.2 & 12.5 \\
GRB 111209A & 0 & 0 & 0 & 0 & 0 & 0 & 0 & 0 & 10.7 & 85.7 & 3.6 & 0 & 0 \\
GRB 101225A & 64.4 & 0 & 0 & 0 & 0 & 0 & 0 & 0 & 5.7 & 21.8 & 2.3 & 3.4 & 2.3 \\
GRB 100316D & 0 & 0 & 0 & 0 & 0 & 0 & 0 & 0 & 13.5 & 34.8 & 11.2 & 9 & 31.5 \\
GRB 090417B & 64 & 0 & 0 & 0 & 0 & 0 & 0 & 0 & 0 & 0 & 9 & 14.6 & 12.4 \\
GRB 060218 & 44 & 0 & 0 & 0 & 0 & 0 & 0 & 0 & 2.4 & 10.7 & 2.4 & 6 & 34.5 \\ \hline
\end{tabular}
\caption{First detection timescale of the ulGRBs by the count-rate trigger prototype software within the fully coded field of view (percentage of simulated burst). \label{table:firts-trig-slice-tc-cnt}}
\end{center}
\end{table}

Both triggers seem to be able to detect the ulGRBs well. 
However, the fully coded field of view represents a small fraction of the total field of view.
We use the cosine of angle $\theta$ to express the distance between the position of the burst in the sky and the optical axis ($\cos{\theta} = 1$ corresponds to an on-axis position). Since an isotropic source distribution is represented as a flat histogram in equal bins of $\cos{\theta}$, we present the results of our simulated positions in bins of $\cos{\theta}$.
Two effects appear, related to the decrease of the sensitivity towards the sides and the corners of the field of view.
First, the fraction of detections decreases when $\theta$ increases.
Second, as $\theta$ increases, the bursts are detected over longer timescales.
The positions of the events follow an isotropic distribution in the sky, which corresponds to a uniform distribution in $\cos{\theta}$.
From the fraction of detected bursts we can then deduce a detection efficiency.

\begin{figure}[h!]
\centering
\includegraphics[width=.9\textwidth]{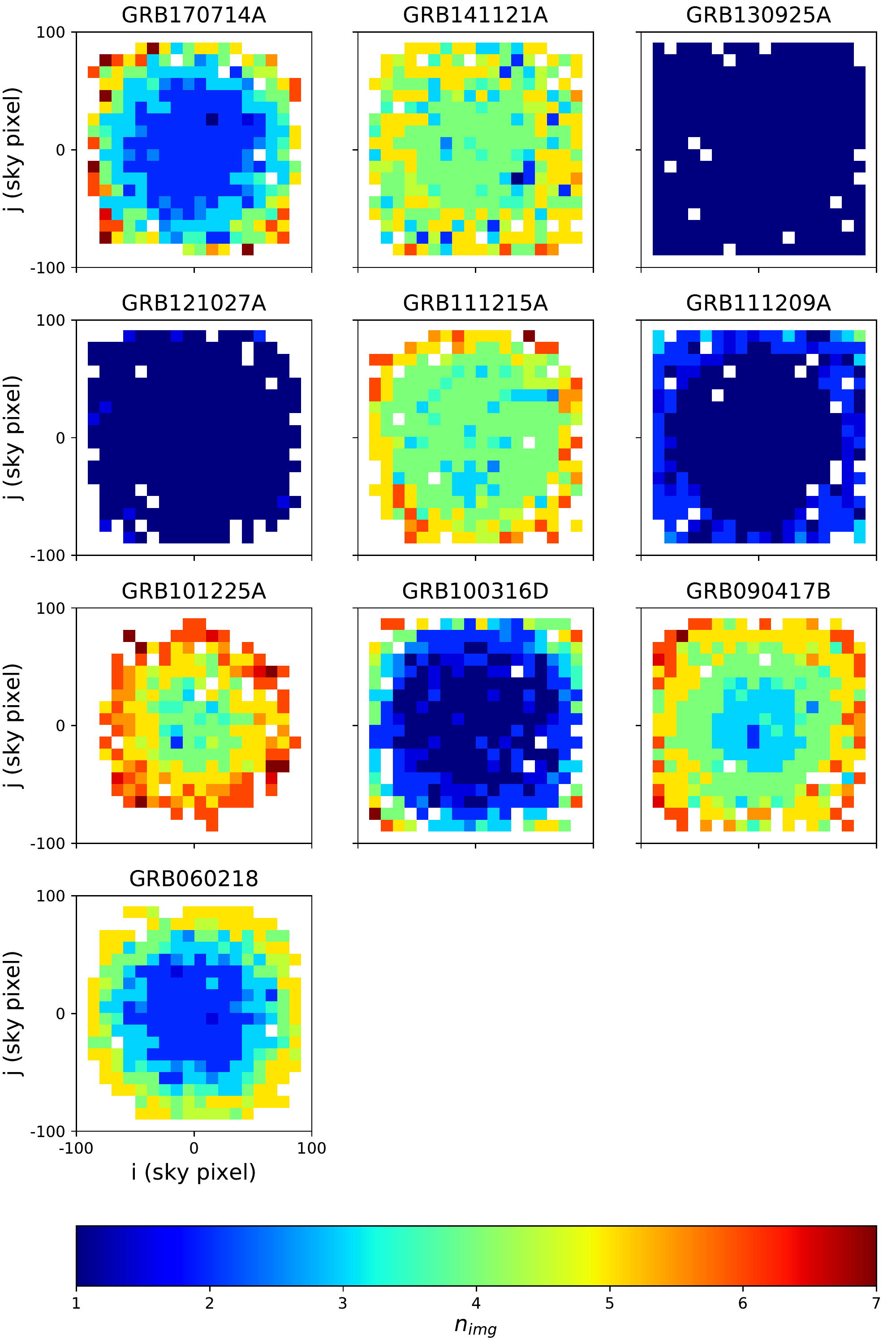}
\caption{Detection of each ulGRB simulated at different positions in the full field of view. The positions are binned by taking the mean in cells of 10x10 sky pixels. The colour gives the image trigger timescale of the first detection. White cells correspond to parts of the sky were bursts are either not injected or not detected. The duration of the timescales indexed $n$ is $2^{n-1} \times 20.48$ s.}
\label{fig:all_Grb}
\end{figure}

\begin{figure}[h!]
\centering
\includegraphics[width=.9\textwidth]{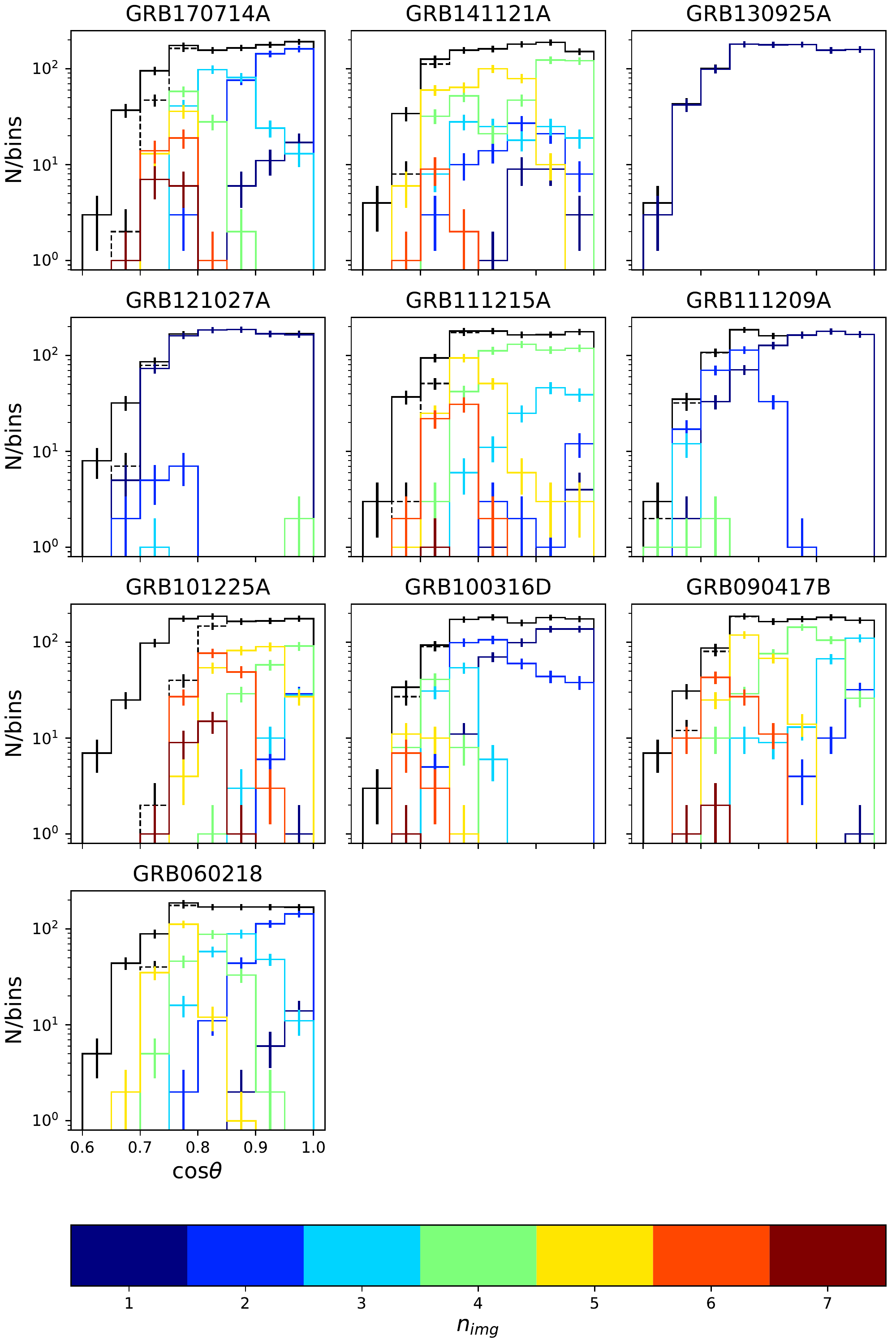}
\caption{Histogram of the detections of each ulGRB simulated at different positions in the total field of view according to $\cos{\theta}$. The line colours give the image trigger timescale of the first detection. The black line shows the 1000 injected positions. The dashed black line shows positions with a detection (for all timescales). The duration of the timescales indexed $n$ is $2^{n-1} \times 20.48$ s.}
\label{fig:theta_all_Grb}
\end{figure}

Figures \ref{fig:all_Grb} and \ref{fig:theta_all_Grb} show both effects for each ulGRB. 
As $\theta$ decreases, the ratio between the number of positions at which the burst is detected and the number of simulated positions decreases.  
Also, for off-axis positions, the first timescale of 20.48 s becomes less efficient and longer timescales are needed to detect the burst. 
However, the loss of off-axis efficiency can be at least partially compensated with a longer integration time.

\begin{figure}[h]
\centering
\includegraphics[width=.8\textwidth]{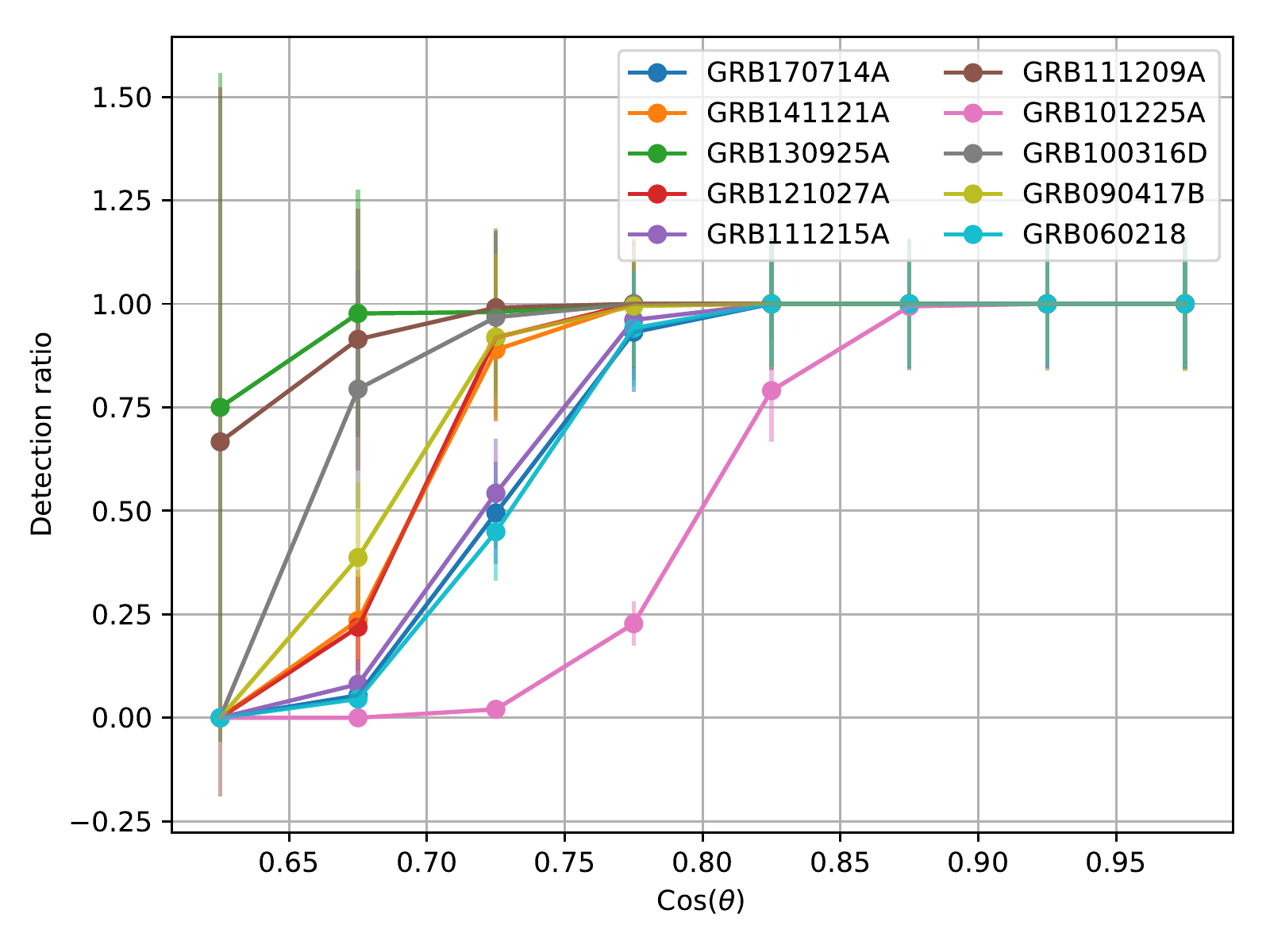}
\caption{Detection ratio (the ratio between the number of positions at which the burst is detected and the number of simulated positions) for each of the 10 ulGRBs as a function of $\cos{\theta}$.}
\label{fig:cosTheta_vs_rDet}
\end{figure}

Figure \ref{fig:cosTheta_vs_rDet} shows the detection ratio for all simulated ulGRBs.
It illustrates the first effect described previously.
Below $\cos{\theta} = 0.875$, the bursts are not always detected, even with the longest timescale of 20 min. 

Moreover, Tab.~\ref{table:firts-trig-slice-ffov-img} and Tab.~\ref{table:firts-trig-slice-ffov-cnt} give the number of detections with the image and count-rate trigger timescales for the 1000 simulated positions for each ulGRB.
For the image trigger, the first timescale is still mostly used but the longer timescales allow to enhance the detection efficiency at off-axis positions.
The timescale 6 is less used and the last timescale is mostly not used.

\begin{table}[h]
\begin{center}
\begin{tabular}{c|c|c|c|c|c|c|c|c}
\hline
\multirow{3}{*}{Name} & \multicolumn{8}{c}{Index $n_{\mathrm{img}}$ and duration of the timescale (seconds)} \\ \cline{2-9}
& \multirow{2}{*}{no detect.} & 1 & 2 & 3 & 4 & 5 & 6 & 7 \\ \cline{3-9}
&            & 20.48 & 40.96 & 81.92 & 163.84 & 327.68 & 655.36 & 1310.72 \\ \hline
GRB 170714A & 98 & 34 & \textbf{411} & 257 & 101 & 50 & 35 & 14 \\
GRB 141121A & 44 & 22 & 83 & 124 & \textbf{396} & 319 & 12 & 0 \\
GRB 130925A & 4 & \textbf{996} & 0 & 0 & 0 & 0 & 0 & 0 \\
GRB 121027A & 40 & \textbf{941} & 16 & 1 & 2 & 0 & 0 & 0 \\
GRB 111215A & 87 & 6 & 18 & 127 & \textbf{521} & 183 & 57 & 1 \\
GRB 111209A & 5 & \textbf{741} & 235 & 15 & 4 & 0 & 0 & 0 \\
GRB 101225A & 304 & 1 & 35 & 41 & 179 & \textbf{257} & 157 & 26 \\
GRB 100316D & 13 & \textbf{454} & 352 & 91 & 57 & 22 & 10 & 1 \\
GRB 090417B & 34 & 1 & 46 & 209 & \textbf{389} & 227 & 91 & 3 \\
GRB 060218 & 107 & 22 & \textbf{313} & 222 & 174 & 162 & 0 & 0 \\ \hline
Sum & 736 & \textbf{3218} & 1509 & 1087 & 1823 & 1220 & 362 & 45 \\ \hline
\end{tabular}
\caption{For each ulGRB, distribution of the number of bursts according to its first detection timescale by the image trigger prototype software, out of  1000 simulations within the total field of view. \label{table:firts-trig-slice-ffov-img}}
\end{center}
\end{table}

Also, with the count-rate trigger, longer timescales up to 20.48 s are more frequently used.
However, over all 10000 simulations (for our 10 busts), the count-rate trigger is less efficient to detect ulGRBs (it does not detect 52.26$\%$ of the events whereas the image trigger missed 7.36 $\%$).

\begin{table}[h]
\begin{center}
\begin{tabular}{c|c|c|c|c|c|c|c|c|c|c|c|c|c}
\hline
\multirow{3}{*}{Name} & \multicolumn{13}{c}{Index $n_{\mathrm{cr}}$ and duration of the timescale (milliseconds)} \\ \cline{2-14}
 & \multirow{2}{*}{no detect.} & 1 & 2 & 3 & 4 & 5 & 6 & 7 & 8 & 9 & 10 & 11 & 12 \\ \cline{3-14}
 &            & 10 & 20 & 40 & 80 & 160 & 320 & 640 & 1280 & 2560 & 5120 & 10240 & 20480 \\ \hline
GRB 170714A & 616 & 0 & 0 & 0 & 0 & 0 & 1 & 37 & 130 & \textbf{189} & 21 & 2 & 4 \\
GRB 141121A & 611 & 0 & 0 & 0 & 0 & 0 & 0 & 1 & 15 & 13 & 86 & 103 & \textbf{171} \\
GRB 130925A & 0 & 0 & 0 & 0 & 0 & 0 & 36 & \textbf{820} & 44 & 92 & 7 & 1 & 0 \\
GRB 121027A & 74 & 0 & 0 & 0 & 0 & 0 & 0 & 2 & 21 & 239 & \textbf{368} & 269 & 27 \\
GRB 111215A & 801 & 0 & 0 & 0 & 0 & 0 & 0 & 0 & 4 & 3 & 62 & \textbf{98} & 32 \\
GRB 111209A & 73 & 0 & 0 & 0 & 0 & 0 & 0 & 1 & 18 & 278 & 179 & \textbf{314} & 137 \\
GRB 101225A & 936 & 0 & 0 & 0 & 0 & 0 & 0 & 0 & 10 & \textbf{37} & 4 & 3 & 10 \\
GRB 100316D & 338 & 0 & 0 & 0 & 0 & 0 & 0 & 3 & 28 & 95 & 52 & 136 & \textbf{348} \\
GRB 090417B & 906 & 0 & 0 & 0 & 0 & 0 & 0 & 0 & 0 & 0 & 11 & 31 & \textbf{52} \\
GRB 060218 & 871 & 0 & 0 & 0 & 0 & 0 & 0 & 0 & 2 & 14 & 6 & 10 & \textbf{97} \\ \hline
Sum & 5226 & 0 & 0 & 0 & 0 & 0 & 37 & 864 & 272 & 960 & 796 & \textbf{967} & 878 \\ \hline
\end{tabular}
\caption{For each ulGRB, distribution of the number of bursts according to its first detection timescale by the count-rate trigger prototype software, out of  1000 simulations within the total field of view. \label{table:firts-trig-slice-ffov-cnt}}
\end{center}
\end{table}

Together with the ulGRBs lightcurve, Fig. \ref{fig:lc_ulGRB} shows the cumulative distributions of the time of detection for both triggers (image in blue, count-rate in green).

Following this, we conclude that the image trigger is rather well adapted to detect ulGRBs.
The results are strongly dependent on the hypothesis that there are neither the Earth nor known sources in the field of view.
However, the results give us confidence in ECLAIRs' ability to detect ulGRBs.
ECLAIRs would have been able to detect the ulGRBs detected by BAT so far in our sample, with a efficiency of 100$\%$ in the fully coded field of view and a loss of 7.36 $\%$ in the total field of view.

% --------------------------------------------
\subsection{Signal to Noise Ratio}
% --------------------------------------------

In this section we study the Signal to Noise Ratio (SNR) of the ulGRBs detected by the image trigger.
The SNR is computed from the sky image obtained by the deconvolution of the shadowgram. 
The deconvolution leads to sky images in counts (C) and in variance (V). 
The SNR is defined for each pixel $i$ of the sky image by $\mathrm{SNR}_i = C_i/\sqrt{V_i}$. 
The detection threshold is SNR=6.5.

\begin{figure}[h!]
\centering
\includegraphics[width=.9\textwidth]{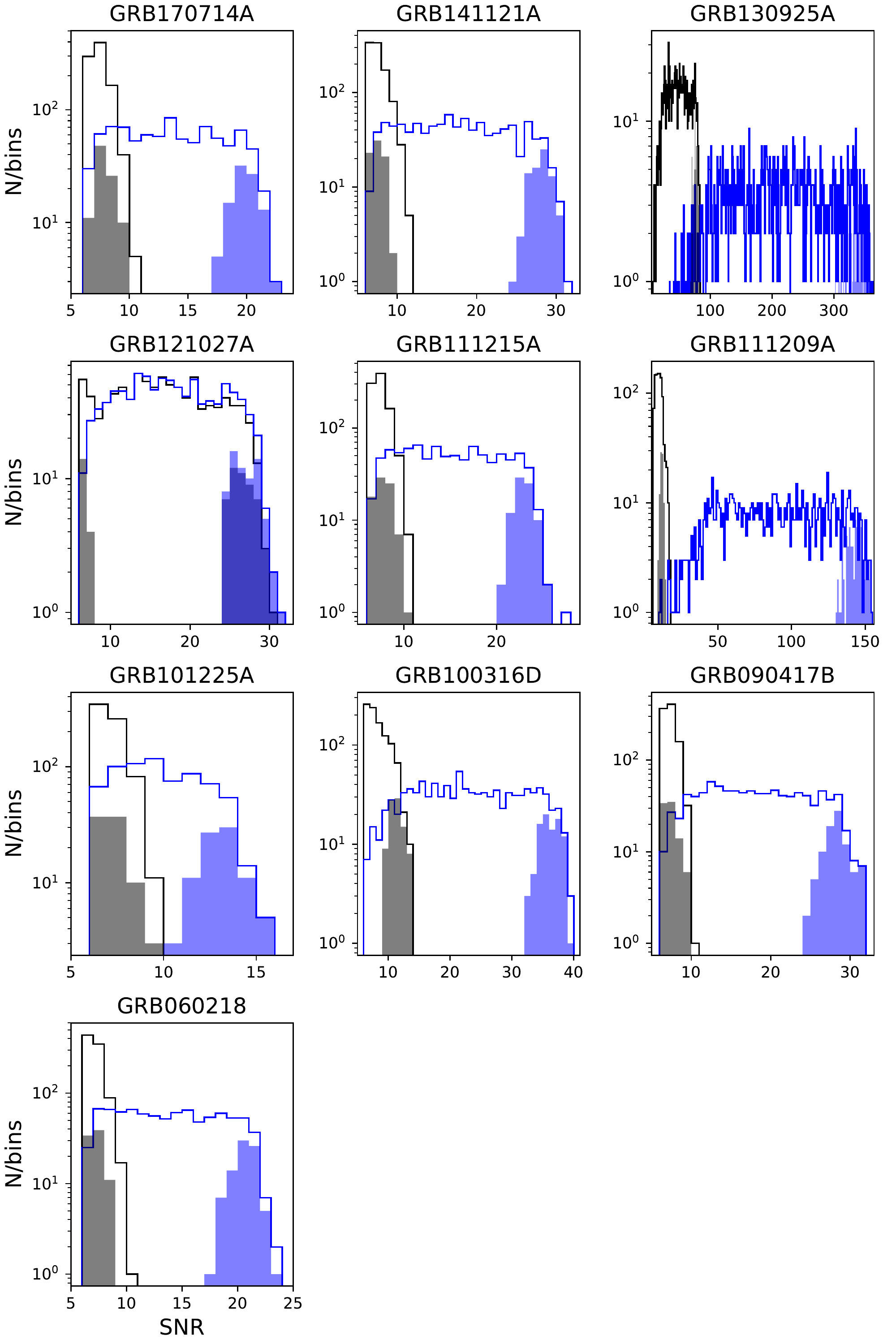}
\caption{SNR of each ulGRB detected by the image trigger at different positions in the total field of view. The black histograms show the first SNR larger than the threshold. The blue histograms show the best SNR for the total duration of the simulation. Filled histograms show values for burst in the fully coded field of view only.}
\label{fig:all_Grb_hist_Snr}
\end{figure}

Figure \ref{fig:all_Grb_hist_Snr} shows the SNR of each ulGRB detected by the image trigger at different positions.
The Tab.~\ref{table:snr} gives the mean SNR values.
We called ``first SNR'' the SNR that first exceeds the 6.5 threshold and ``best SNR'' the best value over the total simulation. 
For most of the ulGRBs, the first SNR is close to the threshold but for the bright bursts like GRB 130925 it reaches larger values.
In Fig. \ref{fig:all_Grb_hist_Snr} we show for the best SNR the influence of the position on the sky.
The filled blue histograms mimic the right part of the blue solid border histograms.
This corresponds to the detection within the fully coded field of view where the sensitivity is the best and leads to higher SNR than in the partially coded field of view.
Also, as the sensitivity is nearly constant in the fully coded field of view, the dispersion of the SNR is smaller than in the total field.
In the partially coded field of view, the best SNR is roughly uniformly distributed between detection limit and maximum SNR in the central field of view. 

\begin{table}[h]
\begin{center}
\begin{tabular}{c|c|c|c|c}
\hline
\multirow{2}{*}{Name} & \multicolumn{2}{c|}{Total field of view} & \multicolumn{2}{c}{Fully coded field of view} \\ \cline{2-5}
 & First & Best & First & Best \\ \hline
GRB 170714A & 7.5 $\pm$ 0.8 & 13.7 $\pm$ 4.2 & 7.9 $\pm$ 0.7 & 19.9 $\pm$ 1.1 \\
GRB 141121A & 7.6 $\pm$ 1.0 & 17.8 $\pm$ 6.4 & 7.5 $\pm$ 0.7 & 28.1 $\pm$ 1.3 \\
GRB 130925A & 47.9 $\pm$ 18.4 & 212.9 $\pm$ 80.5 & 76.6 $\pm$ 3.6 & 337.0 $\pm$ 15.1 \\
GRB 121027A & 16.7 $\pm$ 6.1 & 17.7 $\pm$ 6.0 & 21.4 $\pm$ 8.8 & 27.0 $\pm$ 1.6 \\
GRB 111215A & 7.5 $\pm$ 0.8 & 15.2 $\pm$ 5.0 & 7.8 $\pm$ 0.8 & 22.9 $\pm$ 1.0 \\
GRB 111209A & 10.0 $\pm$ 2.3 & 89.6 $\pm$ 35.1 & 12.0 $\pm$ 1.0 & 143.2 $\pm$ 6.0 \\
GRB 101225A & 7.2 $\pm$ 0.6 & 9.9 $\pm$ 2.2 & 7.3 $\pm$ 0.7 & 13.1 $\pm$ 1.1 \\
GRB 100316D & 8.4 $\pm$ 1.7 & 23.1 $\pm$ 8.5 & 11.3 $\pm$ 1.1 & 36.1 $\pm$ 1.6 \\
GRB 090417B & 7.4 $\pm$ 0.7 & 18.4 $\pm$ 6.4 & 7.5 $\pm$ 0.8 & 28.3 $\pm$ 1.5 \\
GRB 060218 & 7.2 $\pm$ 0.6 & 14.0 $\pm$ 4.4 & 7.2 $\pm$ 0.5 & 20.6 $\pm$ 1.1 \\ \hline
\end{tabular}
\caption{First and best mean SNR for each ulGRB in the total and the fully coded field of view.\label{table:snr}}
\end{center}
\end{table}

% --------------------------------------------
\subsection{Localisation}
% --------------------------------------------

\begin{figure}[h!]
\centering
\includegraphics[width=.7\textwidth]{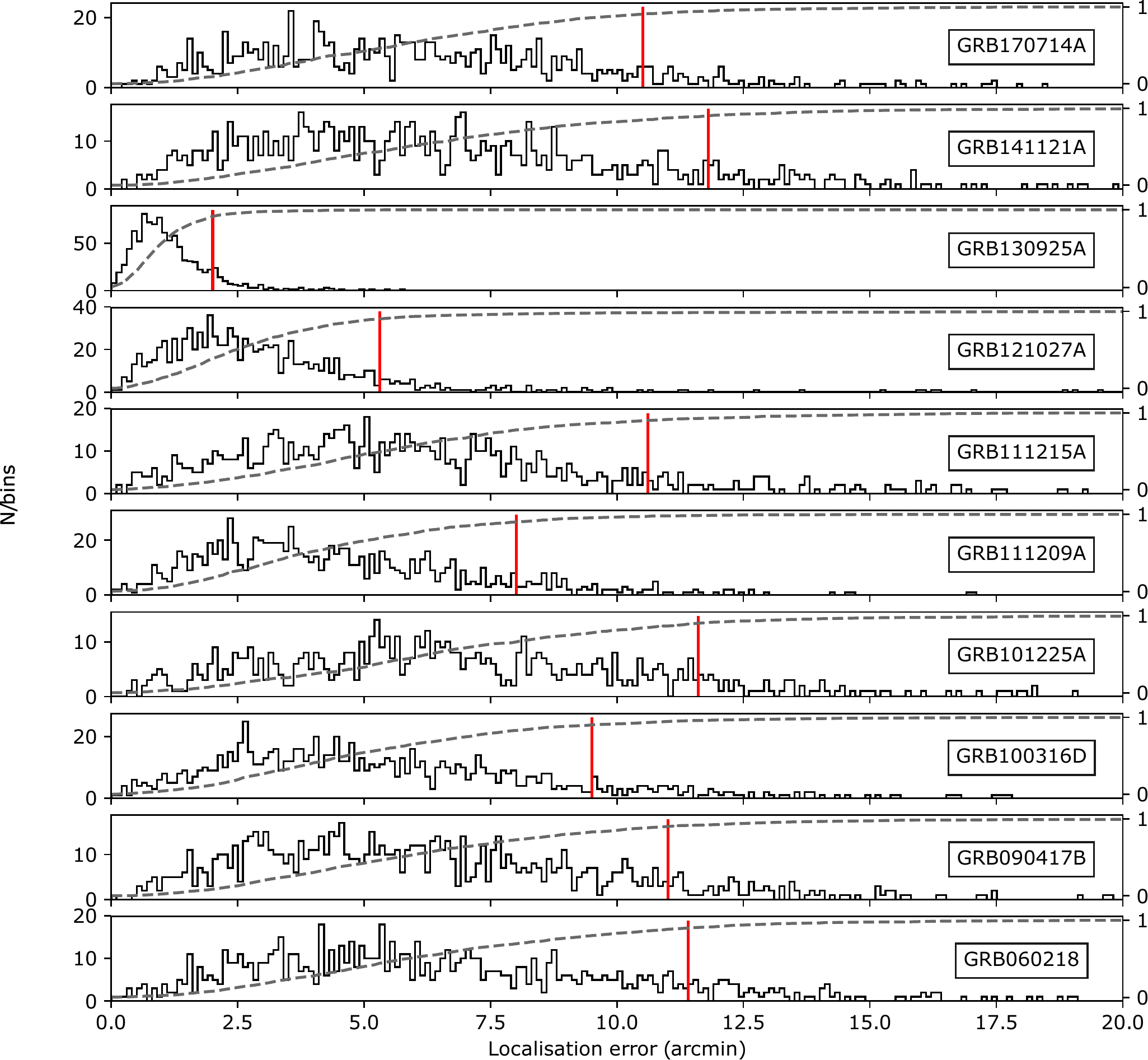}
\caption{Localisation error (in arcmin) for each of the ulGBRs detected by the image trigger in the total field of view. The dashed curve shows the cumulative distribution. The red vertical bar shows the localisation error for which the cumulative distribution reaches a value larger than 90$\%$.}
\label{fig:loc}
\end{figure}

The ulGRB photons are ray-traced from an isotropic random position in the total field of view. 
During both trigger executions, sky images are produced and an excess is searched for. 
An excess is found at a given pixel position in the sky image (integer). 
Then, the newly found source is fitted with a gaussian and a more accurate position from the centre of the gaussian is returned.
We are interested in the localisation accuracy, i.e. the distance on the sky between the injected position and the fitted position.
We know the true position of the burst and the fitted detected position on the sky with coordinates expressed in the detector frame.
We then convert both positions in the equatorial frame and compute the separation in arcminutes.  
Figure \ref{fig:loc} shows the localisation error for each of the ulGRBs detected by the image trigger. 
Bursts from the fully coded field of view are well localised with an error less than 12 arcmin which is the typical ECLAIRs localisation error.
The localisation will allow to quickly identify the position of the burst in order to observe its afterglow with MXT (64$\times$64 arcmin squared field of view), VT (26$\times$26 arcmin squared field of view) and ground instruments.

\begin{figure}[h!]
\centering
\includegraphics[width=.9\textwidth]{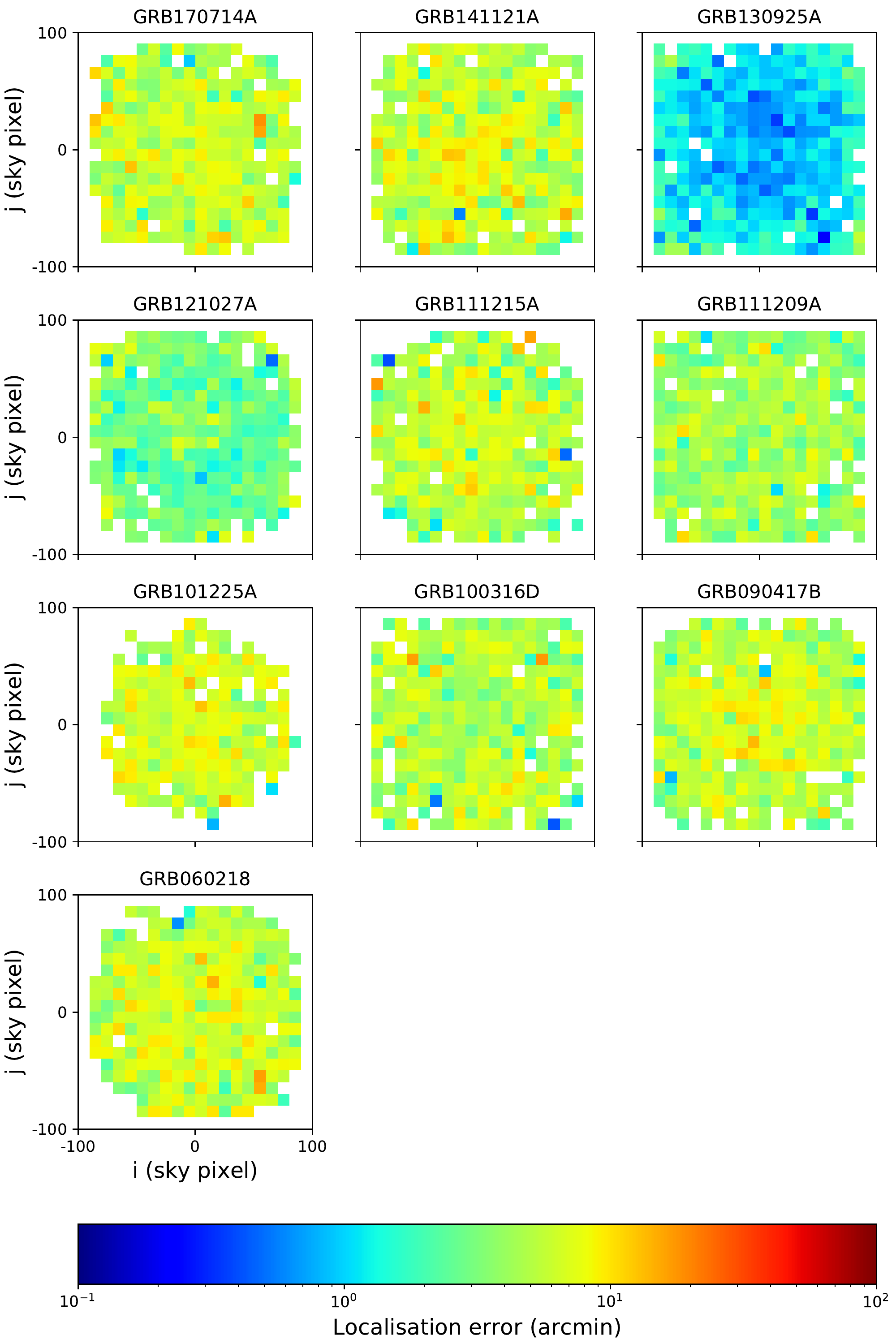}
\caption{Localisation error (arcmin) for each of the ulGRB according to the simulated position in the sky (pixel). The positions are binned by taking the mean in cells of 10$\times$10 sky pixels. White cells correspond to parts of the sky were the bursts are either not injected or not detected.}
\label{fig:all_Grb_loc}
\end{figure}

In the total field of view, the fraction of the simulations leading to a localisation error larger than 12 arcmin is less than 10$\%$ among all the ulGRBs.
Figure \ref{fig:all_Grb_loc} shows the localisation error for all the ulGRB injected at different positions in the field of view, except close to the edges (we remind that 10 pixels have been discarded at the edges due to computation issues that will be solved in the final version of the trigger software).
The localisation errors shown are statistical only and do not take into account systematic effects such as mask-detector misalignment etc.
Because of its brightness, GRB 130925A has a very good localisation (the cumulative distribution reaches a value larger than 90$\%$ for an error of 2.01 arcmin in the total field of view).
This is precisely the kind of phenomenon that requires fast multi-wavelength follow-up.
On the opposite, fainter bursts such as GRB 060218 are less well localised (the cumulative distribution reaches a value larger than 90$\%$ for an error of 11.41 arcmin in the total field of view).

% --------------------------------------------
\subsection{Detection energy strip}
% --------------------------------------------

We also investigate the so-called ``energy strips'' on which the bursts were first detected.
The triggers are working on 4 energy strips: 4-120, 4-50, 4-20 and 20-120 keV configured in our prototype trigger in order to be sensitive to various kinds of bursts (X-ray rich and standard ones).
Results are shown in Tab.~\ref{tab:iestrip}.
For the image trigger, all the ulGRBs are mainly detected on the broadest energy strip. 
Only $\sim 6\%$ of the injected bursts lead to a detection on a smaller strip.
With the count-rate trigger, the 4-120 keV strip is still preferred but some detections occur in the 4-50 keV strip ($\sim 13\%$) and few in 4-20 keV ($\sim 3.8\%$).
Also, with the count-rate trigger, few bursts are detected on the hard-energy strip of 20-120 keV.

\begin{table}[h]
\begin{center}
\begin{tabular}{c|c|c|c|c||c|c|c|c}
\hline
\multirow{2}{*}{Name} & \multicolumn{4}{c}{image} & \multicolumn{4}{c}{count-rate} \\ \cline{2-9}
 & 4-120 & 4-50 & 4-20 & 20-120 & 4-120 & 4-50 & 4-20 & 20-120 \\ \hline \hline
GRB 170714A & 869 & 32 & 1 & 0 & 298 & 84 & 2 & 0 \\
GRB 141121A & 909 & 45 & 2 & 0 & 211 & 146 & 32 & 0 \\
GRB 130925A & 996 & 0 & 0 & 0 & 685 & 303 & 12 & 0 \\
GRB 121027A & 948 & 11 & 1 & 0 & 585 & 302 & 39 & 0 \\
GRB 111215A & 877 & 35 & 1 & 0 & 142 & 52 & 5 & 0 \\
GRB 111209A & 992 & 3 & 0 & 0 & 790 & 115 & 8 & 14 \\
GRB 101225A & 609 & 68 & 19 & 0 & 28 & 34 & 2 & 0 \\
GRB 100316D & 871 & 80 & 36 & 0 & 197 & 219 & 246 & 0 \\
GRB 090417B & 891 & 65 & 10 & 0 & 45 & 40 & 8 & 1 \\
GRB 060218 & 713 & 132 & 48 & 0 & 41 & 56 & 32 & 0 \\ \hline
Sum & 8675 & 471 & 118 & 0 & 3022 & 1351 & 386 & 15 \\ \hline
\end{tabular}
\end{center}
\caption{For each ulGRB, distribution of the number of bursts according to its best energy strip detected in the first timescale by the image trigger and count-rate prototype softwares, out of 1000 simulations within the total field of view (in this table the energy strip with the best SNR value takes precedence over the others, therefore a 0 in a column does not mean the burst is not detected in the corresponding energy strip). \label{tab:iestrip}}
\end{table}

% --------------------------------------------
\subsection{Conclusions on the detectability of the ulGRB sample with ECLAIRs}
% --------------------------------------------

From the results presented in this section, we conclude that ECLAIRs will detect ulGRBs and achieve a good efficiency for Swift-like ulGRBs.

\begin{figure}[h!]
\centering
\includegraphics[width=.9\textwidth]{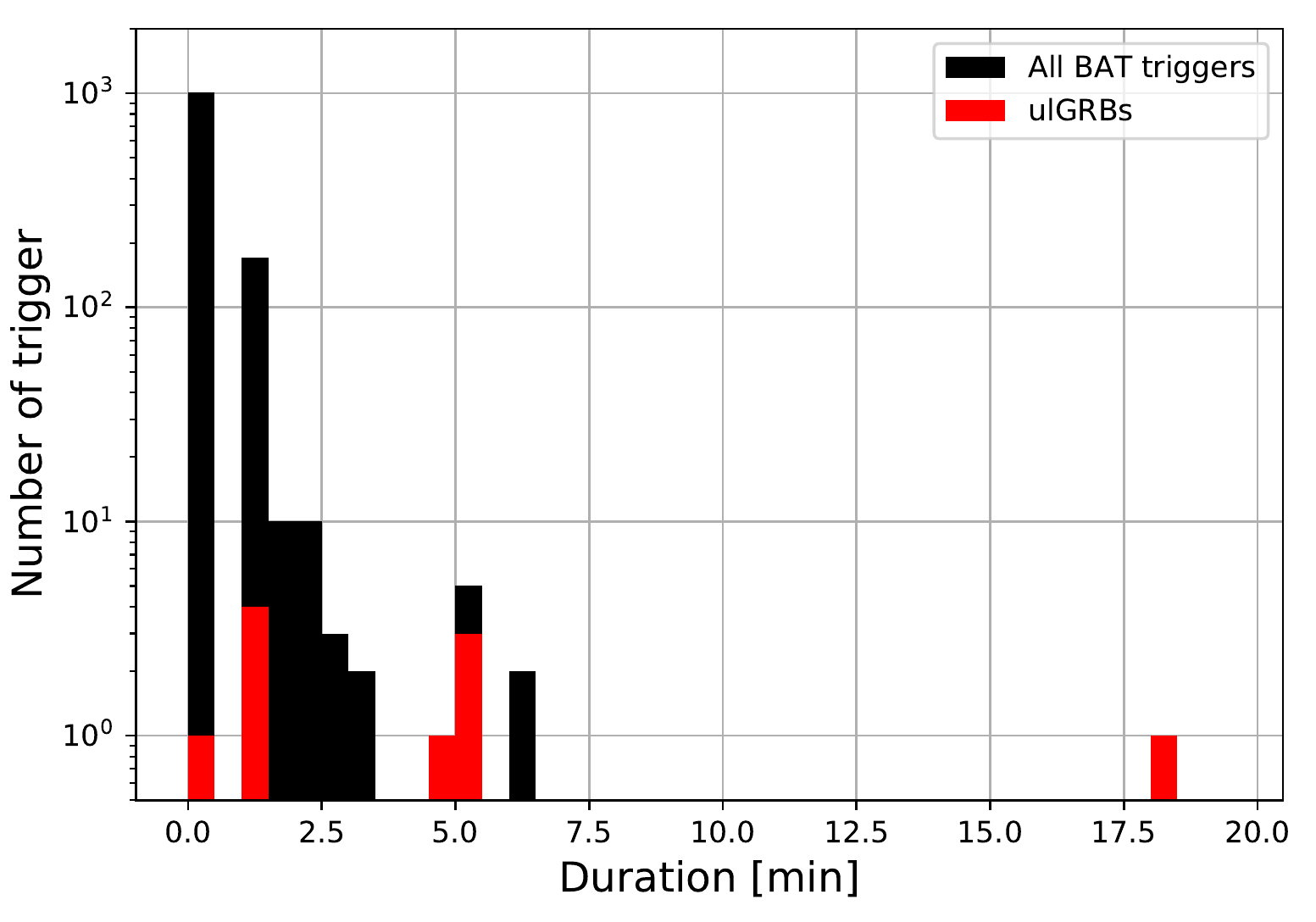}
\caption{Distribution of the BAT trigger timescales (data from \url{https://swift.gsfc.nasa.gov/results/batgrbcat/}). The longest scale of $\sim$ 18 min was used to detect the ulGRB 101225A.}
\label{fig:swift_triggers}
\end{figure}

In this section, we have shown that, among all the timescales of the image trigger, the ones with exposures longer than $\approx$ 10 min ($n_{\mathrm{img}} \geq 6$) are less used than the others, even for bursts at off-axis positions. This behaviour is also observed for BAT trigger that mainly used timescales shorter than 5 min (see Fig. \ref{fig:swift_triggers}). 

Longer timescales up to 20 min may be useful to enhance the detection efficiency for intrinsically fainter ulGRBs, possibly not seen by Swift, or more distant ones. This point will be presented in Sec.~\ref{sec:redshifted}.
 
The onboard triggers are not designed to process images with exposures longer than 20 min, because of built-up of artefacts in very long exposure images, resulting from imperfect background modelling and subtraction onboard. 
But analyses of longer exposures may be achieved on the ground, in particular by the so-called ``off-line searches'' for transient events which may have been missed by the onboard trigger.
Such monitoring will possibly allow to better measure the duration of the burst and to decide whether the burst belongs to the ulGRBs family or not.

% --------------------------------------------
\section{Redshifted ulGRBs}
\label{sec:redshifted}
% --------------------------------------------

As shown in the Tab.~\ref{table:ulGRBs}, the distance of all the ulGRBs used in our work have been measured.
With this information, we can build a sample of artificial redshifted ulGRBs.
To do so, the lightcurves and spectra are transported to redshifts between $z = 1$ and $5$ using a code developed at IAP\footnote{svom.iap.fr} (F. Daigne, M. Bocquier).
This code shifts the photon lightcurve and the spectrum from observations at a redshift $z_0$ in an energy band $(E_1, E_2)$ to a redshift $z$ in an energy band $(E_3, E_4)$.
For each time bin $i$ of the lightcurve, the number of photons $N_{i,34}$ in the energy band $(E_3, E_4)$ at redshift $z$ is given by \citep{antier-farfar_detection_2016}:

\begin{align}
    N_{i,34}(z) = N_{i,12}(z_0)
    \cdot \frac{(1+z)}{(1+z_0)}
    \cdot \frac{D_L^2(z_0)}{D_L^2(z)}
    \cdot \frac{
    \int_{\frac{1+z}{1+z_0}E_3}^{\frac{1+z}{1+z_0}E_4} \mathcal{N}_j(z,E)dE}{\int_{E_2}^{E_1} \mathcal{N}_j(z,E)dE}
\end{align}

with $N_{i,12}$ the number of photons in the energy band $(E_1, E_2)$ at a redshift $z_0$, $D_L$ the luminosity distance and $\mathcal{N}_j$ the spectrum within the time bin $j$:

\begin{align}
    \mathcal{N}_j(z,E) = \frac{(1+z)^3}{(1+z_0)^3}
    \cdot \frac{D_L^2(z_0)}{D_L^2(z)}
    \cdot \mathcal{N}_j(z_0,\frac{1+z}{1+z_0}E).
\end{align}

\begin{figure}[h]
\centering
\includegraphics[width=\textwidth]{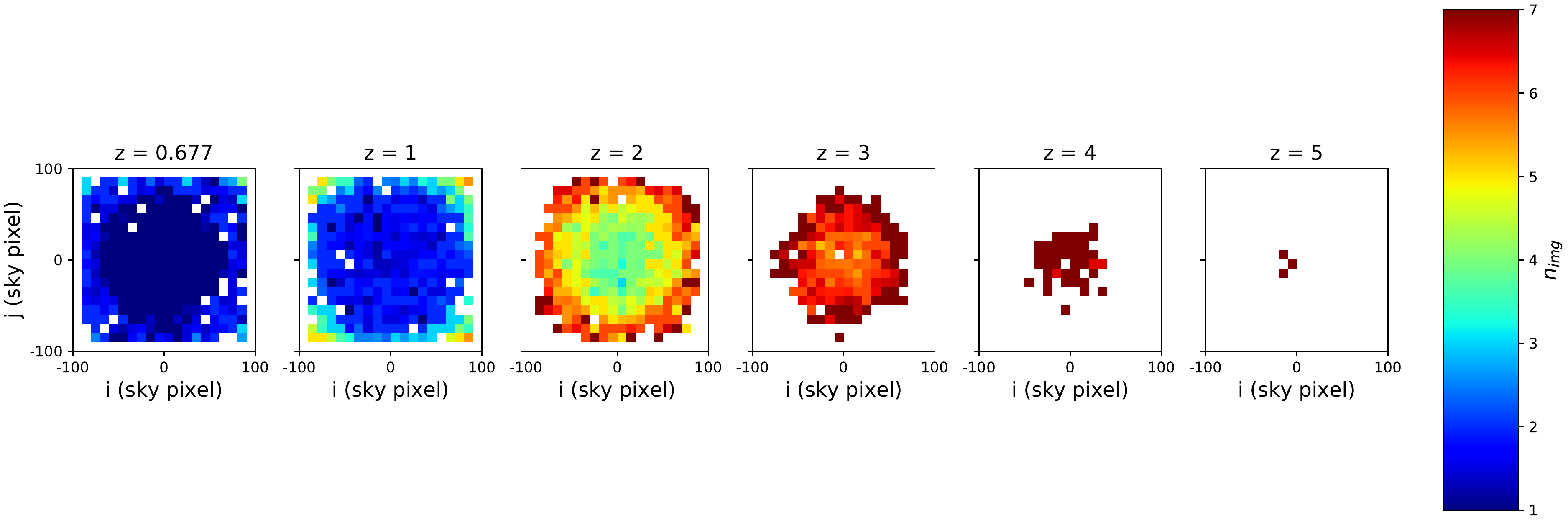}
\caption{Detection of the ulGRB 111209A at different redshifts simulated at different positions in the total field of view. The positions are binned by taking the mean in cells of 10x10 sky pixels. The colour gives the image trigger timescale of the first detection. White cells correspond to parts of the sky were the bursts are either not injected or not detected. The duration of the timescales indexed $n$ is $2^{n-1} \times 20.48$ s.}
\label{fig:z_111209A}
\end{figure}

Once the photon lists at different redshifts have been computed, we can propagate the redshifted ulGRBs photons through ECLAIRs and run the trigger prototype software.
In this section we study the detection of these bursts as a function of the redshift. 
We will focus on the image trigger. 
The redshift has an impact on the detection ratio of the ulGRBs in the total field of view and on the image trigger timescale on which the bursts are detected.
Figure \ref{fig:z_111209A} shows an example for GRB 111209A. 
We can clearly see that the detection ratio decreases with the redshift (as the flux decreases) and that the higher the redshift is, the longer is the timescale of the image trigger to first detect the burst. 
For instance, at $z = 0.677$ the burst has a detection ratio of 1 and is mainly detected thanks to the first timescale of 20.48 s whereas at $z \geq 4$ the detection ratio decreases and the burst is mainly detected within the 20 min timescale. 
Here, we can also observe the effect of the position in the field of view where the time needed to detect the burst increases with $\theta$.

\begin{figure}[h]
\centering
\includegraphics[width=\textwidth]{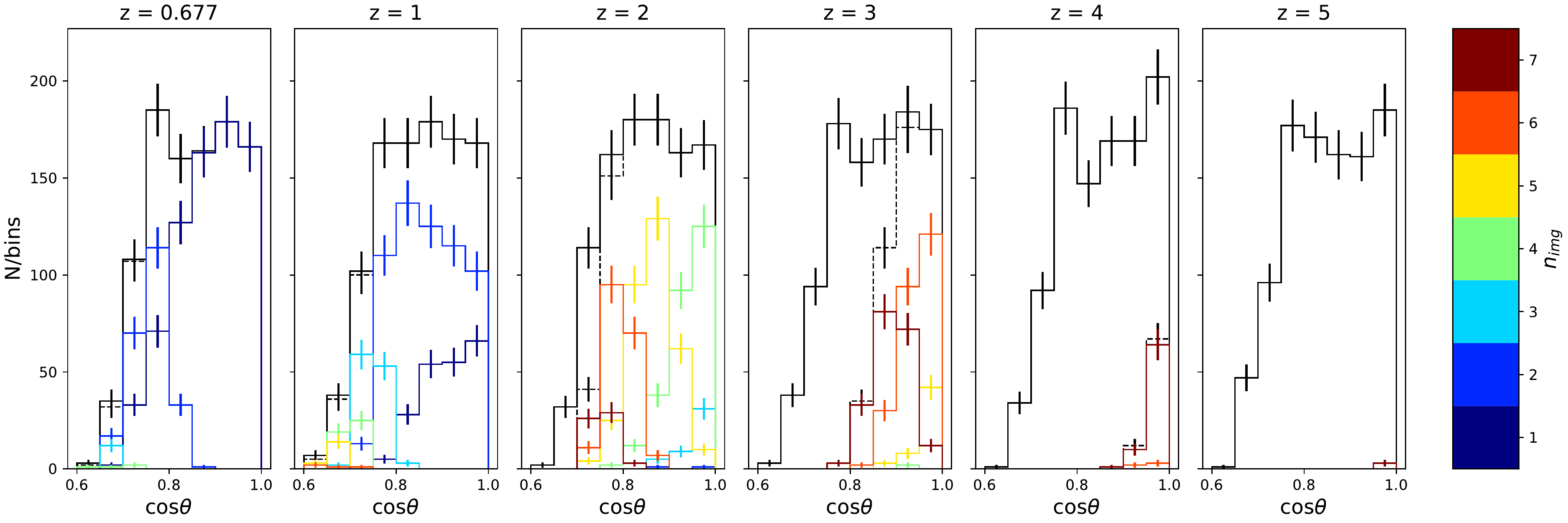}
\caption{Histograms of the detection of the ulGRB 111209A at different redshifts simulated at different positions in the full field of view according to $\cos{\theta}$. The colour lines give the image trigger timescale of the first detection. The black line shows the 1000 injected positions. The dashed black line shows positions with a detection (for all timescales). The duration of the timescales indexed $n$ is $2^{n-1} \times 20.48$ s.}
\label{fig:z_theta_GRB111209A}
\end{figure}

Figure \ref{fig:z_theta_GRB111209A} shows the same kind of histograms as Fig. \ref{fig:theta_all_Grb} for the different redshifts.
As the redshift increases, the shortest timescales become less efficient to the benefit of the longer ones, even at the same position in the field of view.
At very-high redshift too few photons are propagated and the trigger is no longer able to detect the burst, even in the fully coded field of view (for GRB 111209A, the limit is reached for $z > 5$).

\begin{figure}[h]
\centering
\includegraphics[width=\textwidth]{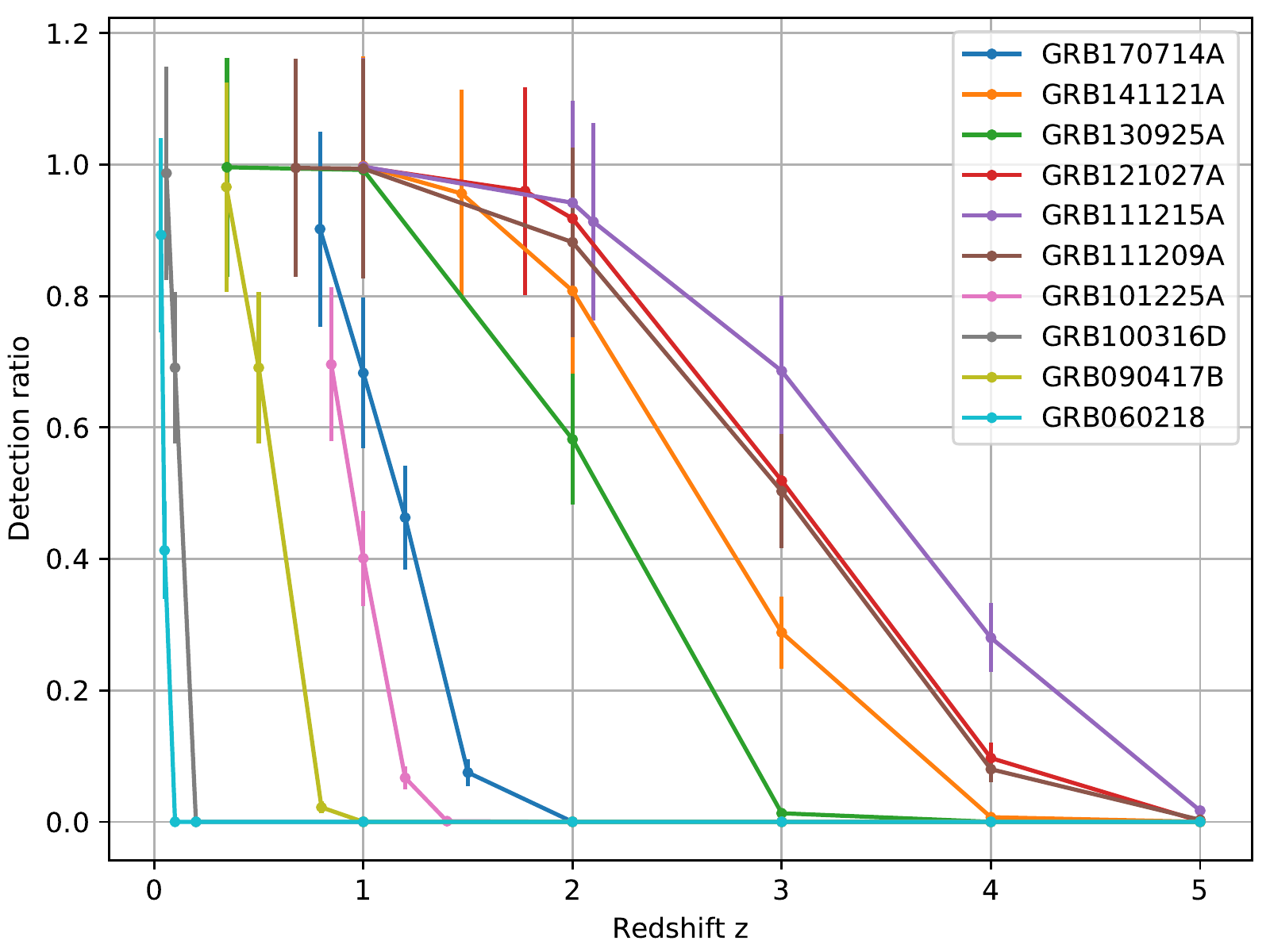}
\caption{Detection fraction (the ratio between the number of positions at which the burst is detected and the number of simulated positions) for each of the 10 ulGRBs according to the redshift.}
\label{fig:z_vs_rDet}
\end{figure}

The same way we did for GRB 111209A, we can compute the detection fraction for each of the 10 ulGRBs.
This is shown by Fig. \ref{fig:z_vs_rDet}.
The first result is that for every ulGRB, the detection fraction decreases with the redshift.
However, the fraction does not decrease the same way for all the bursts depending on their brightness.

% --------------------------------------------
\section{ulGRB rates}
% --------------------------------------------
\label{sec:rates}
In this last section we propose to compute an estimate of the ulGRB rate expected to be detected by ECLAIRs.
As an input we use the table of the 10 ulGRBs detected by Swift in our sample. For each of these bursts, we first estimate the horizon (in redshift) up to which the burst is detectable by Swift and by ECLAIRs in their fully coded field of view. 
Then we compute the estimated number of bursts within each of these horizons, using a population model of GRBs as a function of redshift. 
For each of the Swift bursts, taking the ratio of these numbers, and the fact that Swift has detected one burst, permits to estimate the number of bursts of that kind which will be detected by ECLAIRs. By summation we deduce the total number of ulGRBs of the kind detected by Swift, which we expect to be seen by ECLAIRs.

In order to fairly compare the BAT and ECLAIRs detection capabilities for ulGRBs, we set a limit on the maximum considered ECLAIRs trigger timescale to be equal to the one used by BAT to detect each ulGRB (see Tab.~\ref{table:rate}). Indeed, without such a limit, seeking for the burst with ECLAIRs on its full range of available timescales, without knowing if the burst could possibly have been detected by BAT on a longer timescale, would favour the detection capability of ECLAIRs compared to BAT. Additionally, since the BAT and ECLAIRs trigger timescales do not match exactly, we have performed two sets of calculations for each ulGRB, using the maximum shorter ECLAIRs timescale available called $t_{\mathrm{ECLAIRs,below}}$ (respectively the minimum longer timescale available, called $t_{\mathrm{ECLAIRs,above}}$) compared to the actual trigger timescale of BAT. Also, as GRB 121027A was detected on a very short timescale by BAT (1.024 s), we do not attempt to compute its rate for ECLAIRs, since the ECLAIRs image trigger starts with a smallest timescale of 20.48 s.

As an input, for each of the 9 ulGRBs considered, we use the Swift image trigger information given in the Swift/BAT GRB catalogue\footnote{\url{https://swift.gsfc.nasa.gov/results/batgrbcat/}}, which includes the timescale over which the Swift image trigger has detected the burst ($T_{\mathrm{trig}}$), the signal-to-noise ratio of the detection $\mathrm{SNR}(z_0)$ and the coding fraction $f_{\mathrm{coding}}$.

Considering Swift, in a first step, we determine the horizon of detection by Swift, $z_{h,\mathrm{Swift}}$ for each burst. 
To do so, we generate for each burst redshifted Swift lightcurves, with a similar method as the one used for ECLAIRs in the previous sections (up to $z=5$), considering a BAT sensitive area of 5200~cm$^2$ and without taking into account dead pixels (the same as for ECLAIRs).
We suppose that the background remains the same under the burst signal at each redshift.
For a given burst, we consider each redshifted lightcurve within the image-trigger timescale $T_{\mathrm{trig}}$ used by Swift for the original (non redshifted) burst at $z_0$.
Finally, we determine the number of photons $N_{\mathrm{ph}}(z)$ within $T_{\mathrm{trig}}$ for each burst at each redshift $z$.

The signal-to-noise ratio $\mathrm{SNR_{fcfv}}(z)$ that a burst located in the BAT fully-coded field of view would reach at a given redshift $z$ is then determined by:

\begin{align*}
    \mathrm{SNR_{fcfv}}(z) = \mathrm{SNR}(z_0)\frac{N_{\mathrm{ph}}(z_0)}{N_{\mathrm{ph}}(z)}\frac{1}{f_{\mathrm{coding}}}
\end{align*}

\begin{figure}[h]
\centering
\includegraphics[width=\textwidth]{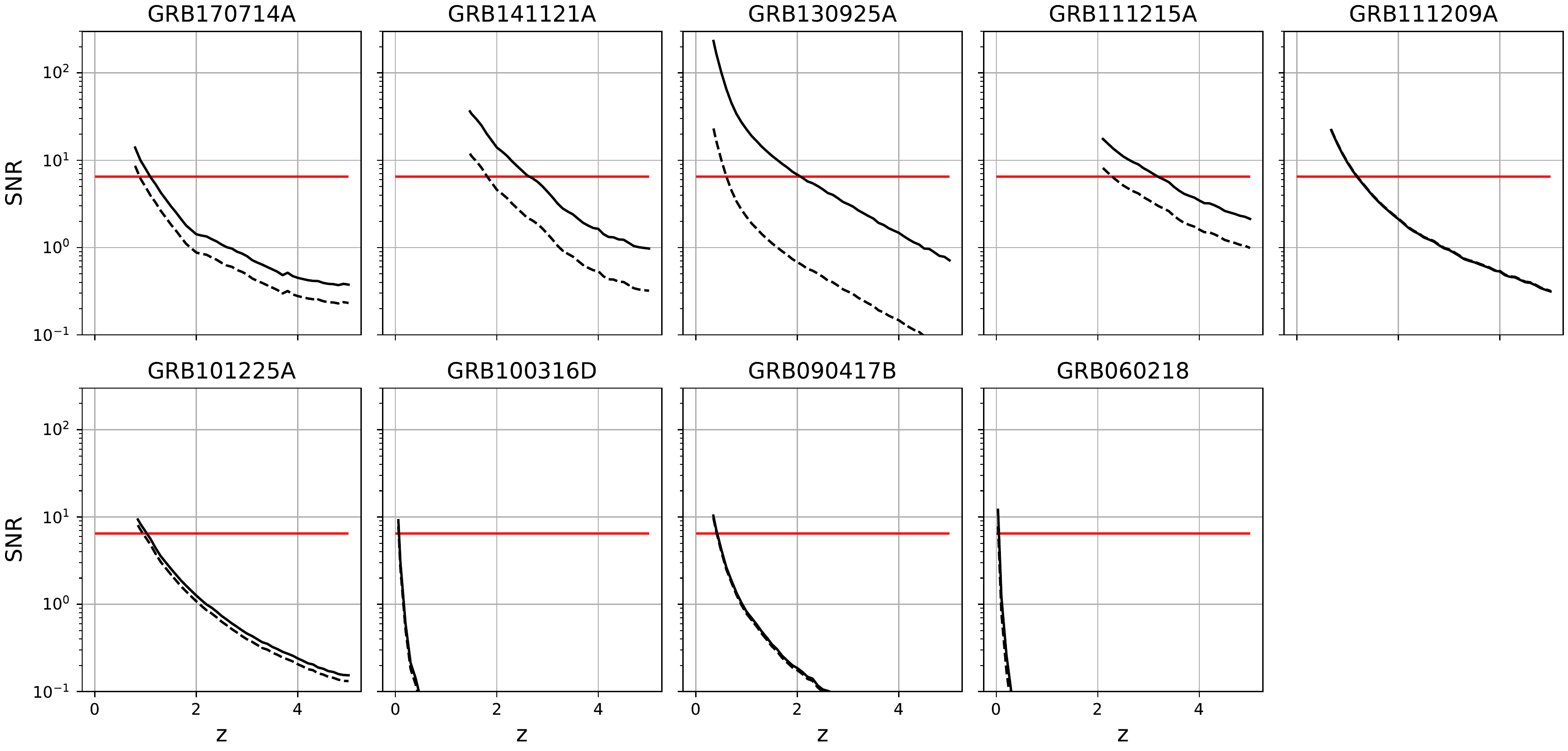}
\caption{For each of the 9 ulGRBs considered, black curve: SNR expected to be measured by Swift/BAT as a function of redshift $z$, the solid line is the SNR expected in the Swift fully coded field of view, whereas the dashed line is the SNR at its true position. The red line indicates a threshold of 6.5.}
\label{fig:all_Grb_swift_snr}
\end{figure}

From the SNR values computed at various redshifts, we determine the horizon $z_{h,\mathrm{Swift}}$ for each burst if it had occurred in the Swift fully-coded field of view, i.e. the redshift value $z_{h,\mathrm{Swift}}$ at which $\mathrm{SNR_{fcfv}}(z_{h,\mathrm{Swift}})$ reaches 6.5.
The results are shown in the black curves of Fig. \ref{fig:all_Grb_swift_snr}. 
For each burst, its horizon $z_{h,\mathrm{Swift}}$ is determined by a linear interpolation of the black curve at the SNR threshold value of 6.5 (red line). Result are listed in Tab.~\ref{table:rate} in the corresponding Swift column. 

In a second step, we compute for each burst the number of bursts in the universe which are closer than its detection horizon by Swift $z_{h,\mathrm{Swift}}$, using the GRB population model determined by the star-formation rate (under the assumption there is no GRB redshift evolution) given by:

\begin{align*}
N(<z_{h,Swift})\propto\int_{0}^{z_{h,\mathrm{Swift}}}\mathrm{SFR}(z)\frac{1}{1+z}\frac{dV(z)}{dz}dz
\end{align*}

with $\frac{dV(z)}{dz}$ the differential comoving volume and $\mathrm{SFR}$ the star-formation rate given by \citep{hopkins_normalization_2006,li_star_2008}:

\begin{align*}
\mathrm{SFR}(z)\propto\frac{0.0157+0.118z}{1+(z/3.23)^{4.66}}.
\end{align*}

The comoving volume is computed under the assumption of a flat $\Lambda$CDM cosmological model ($H_0=70$, $\Lambda=0.7$).

We obtain for each ulGRB the expected (non-normalised) number of bursts $N(<z_{h,\mathrm{Swift}})$ closer than the detection horizon $z_{h,\mathrm{Swift}}$.

Considering ECLAIRs, in order to determine the horizon of detection, we use the same technique as the one applied in the section \ref{sec:redshifted} for the simulation of ECLAIRs GRBs at different redshifts: each burst is redshifted, propagated through the image trigger prototype software, together with background, and the bursts detected above the SNR threshold of 6.5 are considered as triggers. 

\begin{figure}[h]
\centering
\includegraphics[width=\textwidth]{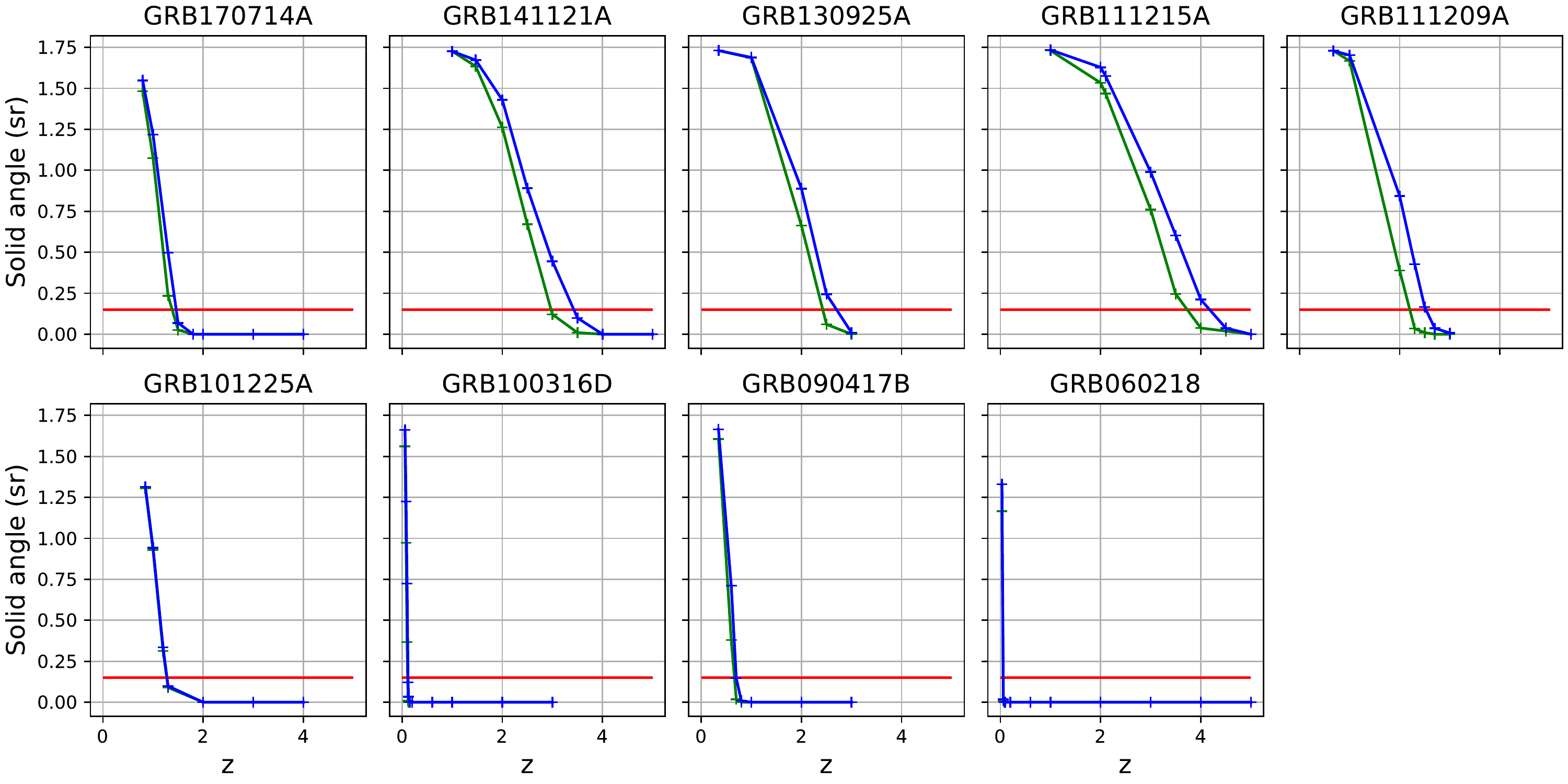}
\caption{For each of the 9 ulGRBs considered, green (blue resp.) curve: size of the field of view (sr) in which the burst is detected considering the timescales shorter than the one just above (below resp.) the BAT one, as a function of the redshift $z$. The red line indicates a threshold of 0.15 sr corresponding to the ECLAIRs fully-coded filed of view.}
\label{fig:all_Grb_ecl_sr}
\end{figure}

For each burst and each redshift, we determine the off-axis solid angle in which the burst is detected by the image trigger, considering two cases for the maximum allowed ECLAIRs timescale durations: those shorter than $t_{\mathrm{ECLAIRs,below}}$ and those shorter than $t_{\mathrm{ECLAIRs,above}}$, bearing in mind that the BAT timescale duration is between these two values.
The solid angle values are shown in Fig. \ref{fig:all_Grb_ecl_sr} as the green curves for the timescale shorter than $t_{\mathrm{ECLAIRs,below}}$ and the blue curves for the timescale shorter than $t_{\mathrm{ECLAIRs,above}}$. 
When the redshift increases, bursts are mainly detected in the centre of the field of view, and the green curve (and blue curve resp.) crosses the limit of the solid angle of 0.15 sr at an endpoint $z_{h,\mathrm{ECLAIRs,below}}$ (and $z_{h,\mathrm{ECLAIRs,above}}$ resp.), the highest redshift at which the GRB is still detected in the fully coded field of view, which is considered as the horizon of detection for ECLAIRs.

For each of the blue and green line, this redshift horizon $z$ lies between two simulated points with redshift $z_{\mathrm{low}}$ and $z_{\mathrm{high}}$. The value of $z$ is determined by the intersection between the red line (at solid angle 0.15 sr) and the straight line joining the points with abscissa $z_{\mathrm{low}}$ and $z_{\mathrm{high}}$. We use $z_{\mathrm{low}}$ and $z_{\mathrm{high}}$ as bounds of the range obtained for $z$ from our limited number of simulations and we consider as systematic error bars of $z$ the values $z$-$z_{\mathrm{high}}$ and $z$-$z_{\mathrm{low}}$ reported in Tab.~\ref{table:rate2}.

The same way as we did for Swift, considering the same GRB population model as a function of redshift, we compute the number of bursts closer than the ECLAIRs redshift horizon $N(<z_{h,\mathrm{ECLAIRs,below}})$ and $N(<z_{h,\mathrm{ECLAIRs,above}})$.

Finally, for each ulGRB, we get numbers $N(<z_{h,\mathrm{Swift}})$, $N(<z_{h,\mathrm{ECLAIRs,below}})$ and $N(<z_{h,\mathrm{ECLAIRs,above}})$ which correspond respectively to the number of bursts closer than the Swift and ECLAIRs detection horizons $z_{h,\mathrm{Swift}}$, $z_{h,\mathrm{ECLAIRs,below}}$ and $z_{h,\mathrm{ECLAIRs,above}}$ in the fully coded field of view. Given the fact that (for each ulGRB) Swift has detected one specimen, the expected detection rate by ECLAIRs (for an ulGRB of the same kind) is $\mathrm{Rate_{\mathrm{below}}} = N_{\mathrm{ECLAIRs,below}}/N_{\mathrm{Swift}}$ and $\mathrm{Rate_{\mathrm{above}}} = N_{\mathrm{ECLAIRs,above}}/N_{\mathrm{Swift}}$. The error range $z_{\mathrm{low}}$ and $z_{\mathrm{high}}$ we obtained on the redshift horizon is propagated to this ratio. The rates obtained are listed in Tab.~\ref{table:rate2}. For example, while Swift has detected one ulGRB of the type GRB 170714 during its 15 years of operation, we can expect between 1.53 and 2.11 GRBs of that kind detected by ECLAIRs during the same (hypothetical) observation time of 15 years and with the same (hypothetical) duty cycle as Swift. 

\begin{table}[h]
\begin{center}
\begin{tabular}{c|c|c|c}
\hline
Name & Timescale (s) & SNR$(z_0)$ & $z_h$ \\ \hline \hline
GRB 170714A & 320 & 8.63 & 1.091 \\
GRB 141121A & 288 & 11.91 & 2.641 \\
GRB 130925A & 64 & 23.15 & 2.063 \\
GRB 121027A & 1.024& 7.88 & 1.813 \\
GRB 111215A & 320 & 8.16 & 3.183 \\
GRB 111209A & 64 & 22.54 & 1.185 \\
GRB 101225A & 1088 & 8.06 & 1.025 \\
GRB 100316D & 64 & 7.83 & 0.077 \\
GRB 090417B & 320 & 9.81 & 0.426 \\
GRB 060218 & 80 & 7.84 & 0.067 \\ \hline
\end{tabular}
\end{center}
\caption{Summary table for Swift:  trigger timescales, SNR, redshift horizon. \label{table:rate}}
\end{table}

\begin{table}[h]
\renewcommand{\arraystretch}{1.5}
\begin{center}
\begin{tabular}{c|c|c|c|c|c}
\hline
Name &  Timescale (s) & $z_{h,\mathrm{below}}$ & $z_{h,\mathrm{above}}$ & $\mathrm{Rate_{\mathrm{below}}}$ & $\mathrm{Rate_{\mathrm{above}}}$ \\ \hline \hline
GRB 170714A & 163.84 - 327.68 & 1.38$_{-0.08}^{+0.12}$ & 1.46$_{-0.16}^{+0.04}$ & 1.76$_{-0.23}^{+0.36}$ & 2.00$_{-0.47}^{+0.12}$ \\
GRB 141121A & 163.84 - 327.68 & 2.97$_{-0.47}^{+0.03}$ & 3.43$_{-0.43}^{+0.07}$ & 1.16$_{-0.23}^{+0.01}$ & 1.32$_{-0.15}^{+0.02}$ \\
GRB 130925A & 40.96 - 81.92 & 2.43$_{-0.43}^{+0.07}$ & 2.70$_{-0.20}^{+0.30}$ & 1.30$_{-0.36}^{+0.06}$ & 1.52$_{-0.15}^{+0.20}$ \\
GRB 111215A & 163.84 - 327.68 & 3.73$_{-0.23}^{+0.27}$ & 4.18$_{-0.18}^{+0.32}$ & 1.13$_{-0.05}^{+0.05}$ & 1.20$_{-0.02}^{+0.04}$ \\
GRB 111209A & 40.96 - 81.92 & 2.20$_{-0.20}^{+0.10}$ & 2.52$_{-0.02}^{+0.18}$ & 3.64$_{-0.56}^{+0.27}$ & 4.49$_{-0.06}^{+0.44}$ \\
GRB 101225A & 655.36 - 1310.72 & 1.27$_{-0.07}^{+0.03}$ & 1.28$_{-0.08}^{+0.02}$ & 1.71$_{-0.22}^{+0.09}$ & 1.72$_{-0.24}^{+0.07}$ \\
GRB 100316D & 40.96 - 81.92 & 0.11$_{-0.01}^{+0.01}$ & 0.12$_{-0.02}^{+0.00}$ & 2.78$_{-1.07}^{+0.73}$ & 3.43$_{-1.71}^{+0.09}$ \\
GRB 090417B & 163.84 - 327.68 & 0.66$_{-0.06}^{+0.04}$ & 0.70$_{-0.10}^{+0.00}$ & 3.75$_{-0.95}^{+0.62}$ & 4.37$_{-1.57}^{+0.01}$ \\
GRB 060218 & 40.96 - 81.92 & 0.06$_{-0.02}^{+0.00}$ & 0.06$_{-0.02}^{+0.00}$ & 0.54$_{-0.36}^{+0.14}$ & 0.56$_{-0.39}^{+0.12}$ \\
 \hline 
\end{tabular}
\end{center}
\caption{Summary table for ECLAIRs: for each trigger timescale (below and above the timescale of Swift), redshift horizon $z$ (with errors derived from $z_{\mathrm{low}}$ and $z_{\mathrm{high}}$) and rates (with errors propagated from $z_{\mathrm{low}}$ and $z_{\mathrm{high}}$). \label{table:rate2}}
\end{table}

At the end, if we consider only ECLAIRs timescales below the BAT trigger timescale, in order to get an estimate of the global ulGRB detection rate by ECLAIRs, we sum all the individual detection rates $\mathrm{Rate_{\mathrm{below}}}$ for each of the 9 ulGRBs considered. 
With 9 ulGRBs detected by Swift/BAT and an estimated sum of $18_{-4}^{+2}$ ulGRBs detected by ECLAIRs over the same observation time, this leads to an estimated global detection ratio of $1.97_{-0.45}^{+0.26}$ times more ulGRBs seen by ECLAIRs than by Swift.

However, the Earth passages in the field of view of ECLAIRs reduce the total field of view of 2.04 sr to 65$\%$ of this value (one year average), and the SAA crossings reduce the duty cycle to 86$\%$ of the observing time, leading to an overall duty cycle of about 56$\%$ for SVOM, which is smaller than the one of Swift (75$\%$). Also ECLAIRs has a slightly smaller total field of view than BAT: 2.04 sr versus 2.3 sr \citep{barthelmy_burst_2005}. To account for these differences we correct the previous rate by a factor 0.66 which gives a final rate of $1.30_{-0.30}^{+0.17}$ ulGRBs expected to be detected by ECLAIRs for each one detected by BAT.

In order to evaluate the influence of the low energy threshold (whose design value is 4 keV, value used in the previous analysis), we repeat the simulations with different sets of energy bands. If we set the low energy threshold to a degraded value of 7 keV (10 keV resp.), the rate of ulGRBs detected by ECLAIRs compared to BAT becomes $1.10_{-0.24}^{+0.16}$ ($0.97_{-0.29}^{+0.18}$ resp.), which shows the benefit of the low-energy threshold of ECLAIRs on the expected ulGRB detection rate.

Concluding our analysis, we expect ECLAIRs to detect at least as many ulGRBs as Swift/BAT. The rate we give can be considered a lower limit, which applies only to the small number of ulGRBs that have been detected by Swift so far and simulated through ECLAIRs. SVOM could eventually detect even more ulGRBs thanks to its specific characteristics (low energy threshold opening space for undetected Swift bursts) and with the help of the ground-based ``off-line trigger'' which will use (although with delay) all photons recorded by the detector. 

% --------------------------------------------
\section{Discussion and conclusion}
% --------------------------------------------

In this paper, we studied the ECLAIRs capabilities to detect ulGRBs using a sample of events detected by Swift/BAT. We showed that ECLAIRs will be efficient to detect ulGRBs, even if they occur in the partially coded field of view or at higher redshifts. 

In addition, we determined that ECLAIRs may detect at least as much ulGRBs as Swift/BAT, given the same observing time and taking into account the differences in duty cycles and fields of view of both instruments.

However, as mentioned in Sec.~\ref{sec:svom}, the simulations used in this work are based on the assumption that the Earth is not crossing the field of view of the instrument.  In reality, the Earth will cross the filed of view almost every orbit, and the satellite track will move through the area of the South Atlantic Anomaly (SAA) of the Earth magnetic field for several subsequent orbits each day. Both effects will reduce the 90 min orbit duty cycle to about 50 min (taken into account in our estimate) but they will also create possible gaps in ulGRB lightcurves, which will prevent the image trigger to reach its full efficiency.

Nevertheless, SVOM will be able to observe these events for a long time thanks to its long stable pointings, up to $\sim 20$ hours (with some gaps due to Earth and SAA crossings). This long time monitoring with all four SVOM space instruments ECLAIRs, GRM, MXT and VT will permit to better clarify the actual ultra-long duration of these events. Additionally with the SVOM ground instruments GWAC and GFT (see Tab.~\ref{table:coverage} for a basic comparison of all SVOM instruments), ulGRB emissions could be monitored in the visible and near-infrared bands from their prompt to afterglow phases over a long time.

\begin{table}[h]
\begin{center}
\begin{tabular}{c|c|c}
\hline
\multicolumn{2}{c}{Instruments}  & Coverage \\ \hline \hline
\multirow{4}{*}{Space} & GRM & 15 keV - 5 MeV \\
 & ECLAIRs & 4 - 150 keV \\
 & MXT & 0.2 - 10 keV \\
 & VT & 400 - 1000 nm \\ \hline
\multirow{3}{*}{Ground} & F-GFT & 400 - 1700 nm \\
 & C-GFT & 400 - 1000 nm \\
 & GWAC & 450 - 900 nm \\
 \hline 
\end{tabular}
\end{center}
\caption{Space and ground instruments of the SVOM mission with their spectral coverage. \label{table:coverage}}
\end{table}

Furthermore ECLAIRs has the specificity that all the counts recorded by the instruments will be sent to the ground, with a delay of about 6 to 12 hours, depending on the availability of the X-band download passages. 
On ground there will be the so-called ``offline trigger'' reanalysing those data (with delay), and it may detect ulGRBs on even longer timescales, by combining for example data from several subsequent orbits while ignoring the periods when the Earth masks portions of the field of view or periods when the spacecraft moves trough the SAA.

However, this study give us the opportunity to raise some questions. BAT has detected GRB130925A with a SNR larger than 20 (compared to other ulGRBs which were detected with SNR around 8) at a redshift of 0.348. If we suppose such events are produced at larger distances in the Universe, BAT should have detected them. We have determined a redshift horizon of 2.063 for this burst, which leads to a comoving volume $\sim$ 58 times larger. Therefore the question remains: why so few ulGRBs of that kind have been detected? Was GRB130925A a chance event?

In summary, based on the reasons enumerated here and the results of our study, we believe that SVOM will play a role to understand the nature of the ulGRB phenomenon, to measure the precise duration of each event and to perform multi-wavelength observations of their prompt and afterglow phases, helping ultimately to unveil the nature of their progenitors and to determine if those events really form a distinct class of GRBs.

\begin{acknowledgements}
ECLAIRs is a cooperation between CNES, CEA and CNRS, with CNES acting as prime contractor. This work is supported by the CEA, the ``IDI 2017'' project of the French ``Investissements d'Avenir'' program, financed by IDEX Paris-Saclay (ANR-11-IDEX-0003-02) and by the UnivEarthS Labex program at Sorbonne Paris Cité (ANR-10-LABX-0023 and ANR-11-IDEX-0005-02).
We would like to thank Frédéric Daigne and Maxime Bocquier from IAP, Paris, for providing the tools to transport GRB light curves to different redshifts, developed during the thesis of Sarah Antier at CEA Paris-Saclay.
We also would like to address our thanks to Bruce Gendre for fruitful discussions about ulGRBs. 
Finally, we acknowledge helpful comments from the anonymous referee.
\end{acknowledgements}

\bibliographystyle{aasjournal}
\bibliography{references} 

\begin{thebibliography}{}
\expandafter\ifx\csname natexlab\endcsname\relax\def\natexlab#1{#1}\fi
\providecommand{\url}[1]{\href{#1}{#1}}
\providecommand{\dodoi}[1]{doi:~\href{http://doi.org/#1}{\nolinkurl{#1}}}
\providecommand{\doeprint}[1]{\href{http://ascl.net/#1}{\nolinkurl{http://ascl.net/#1}}}
\providecommand{\doarXiv}[1]{\href{https://arxiv.org/abs/#1}{\nolinkurl{https://arxiv.org/abs/#1}}}

\bibitem[{{Antier-Farfar}(2016)}]{antier-farfar_detection_2016}
{Antier-Farfar}, S. 2016, {PhD Thesis}, Universit\'e Paris-Saclay.
\newblock \url{https://tel.archives-ouvertes.fr/tel-01456239/document}

\bibitem[{Barthelmy {et~al.}(2005)Barthelmy, Barbier, Cummings, Fenimore,
  Gehrels, Hullinger, Krimm, Markwardt, Palmer, Parsons, Sato, Suzuki,
  Takahashi, Tashiro, \& Tueller}]{barthelmy_burst_2005}
Barthelmy, S.~D., Barbier, L.~M., Cummings, J.~R., {et~al.} 2005, Space Sci.
  Rev., 120, \dodoi{10.1007/s11214-005-5096-3}

\bibitem[{Bo{\"e}r {et~al.}(2015)Bo{\"e}r, Gendre, \& Stratta}]{boer_are_2015}
Bo{\"e}r, M., Gendre, B., \& Stratta, G. 2015, ApJ, 800,
  \dodoi{10.1088/0004-637X/800/1/16}

\bibitem[{Campana {et~al.}(2006)Campana, Mangano, Blustin, Brown, Burrows,
  Chincarini, Cummings, Cusumano, Valle, Malesani, M{\'e}sz{\'a}ros, Nousek,
  Page, Sakamoto, Waxman, Zhang, Dai, Gehrels, Immler, Marshall, Mason,
  Moretti, O'Brien, Osborne, Page, Romano, Roming, Tagliaferri, Cominsky,
  Giommi, Godet, Kennea, Krimm, Angelini, Barthelmy, Boyd, Palmer, Wells, \&
  White}]{campana_association_2006}
Campana, S., Mangano, V., Blustin, A.~J., {et~al.} 2006, Nat, 442,
  \dodoi{10.1038/nature04892}

\bibitem[{Cobb {et~al.}(2010)Cobb, Bloom, Perley, Morgan, Cenko, \&
  Filippenko}]{cobb_discovery_2010}
Cobb, B.~E., Bloom, J.~S., Perley, D.~A., {et~al.} 2010, ApJ, 718,
  \dodoi{10.1088/2041-8205/718/2/L150}

\bibitem[{Cucchiara {et~al.}(2015)Cucchiara, Veres, Corsi, Cenko, Perley, Lien,
  Marshall, Pagani, Toy, Capone, Frail, Horesh, Modjaz, Butler, Littlejohns,
  Watson, Kutyrev, Lee, Richer, Klein, Fox, Prochaska, Bloom, Troja,
  {Ramirez-Ruiz}, {de Diego}, Georgiev, Gonz{\'a}lez,
  {Rom{\'a}n-Z{\'u}{\~n}iga}, Gehrels, \& Moseley}]{cucchiara_happy_2015}
Cucchiara, A., Veres, P., Corsi, A., {et~al.} 2015, ApJ, 812,
  \dodoi{10.1088/0004-637X/812/2/122}

\bibitem[{Dezalay {et~al.}(1996)Dezalay, Lestrade, Barat, Talon, Sunyaev,
  Terekhov, \& Kuznetsov}]{dezalay_hardness-duration_1996}
Dezalay, J.~P., Lestrade, J.~P., Barat, C., {et~al.} 1996, ApJL, 471,
  \dodoi{10.1086/310321}

\bibitem[{Evans {et~al.}(2014)Evans, Willingale, Osborne, O'Brien, Tanvir,
  Frederiks, Pal'shin, Svinkin, Lien, Cummings, Xiong, Zhang, G{\"o}tz,
  Savchenko, Negoro, Nakahira, Suzuki, Wiersema, Starling, {Castro-Tirado},
  Beardmore, {S{\'a}nchez-Ram{\'i}rez}, Gorosabel, Jeong, Kennea, Burrows, \&
  Gehrels}]{evans_grb130925a:_2014}
Evans, P.~A., Willingale, R., Osborne, J.~P., {et~al.} 2014, MNRAS, 444,
  \dodoi{10.1093/mnras/stu1459}

\bibitem[{Gendre {et~al.}(2019)Gendre, Joyce, Orange, Stratta, Atteia, \&
  Bo{\"e}r}]{gendre_can_2019}
Gendre, B., Joyce, Q.~T., Orange, N.~B., {et~al.} 2019, MNRAS, 486,
  \dodoi{10.1093/mnras/stz1036}

\bibitem[{Gendre {et~al.}(2013)Gendre, Stratta, Atteia, Basa, Bo{\"e}r, Coward,
  Cutini, D'Elia, Howell, Klotz, \& Piro}]{gendre_ultra-long_2013}
Gendre, B., Stratta, G., Atteia, J.~L., {et~al.} 2013, ApJ, 766,
  \dodoi{10.1088/0004-637X/766/1/30}

\bibitem[{Godet {et~al.}(2014)Godet, Nasser, Atteia, Cordier, Mandrou, Barret,
  Triou, Pons, Amoros, Bordon, Gevin, Gonzalez, G{\"o}tz, Gros, Houret,
  Lachaud, Lacombe, Marty, Mercier, Rambaud, Ramon, Rouaix, Schanne, \&
  Waegebaert}]{takahashi_x-gamma-ray_2014}
Godet, O., Nasser, G., Atteia, J.-., {et~al.} 2014, in {{SPIE Astronomical
  Telescopes}} + {{Instrumentation}}, {Montr\'eal, Quebec, Canada},
  \dodoi{10.1117/12.2055507}

\bibitem[{Goldwurm {et~al.}(2003)Goldwurm, David, Foschini, Gros, Laurent,
  Sauvageon, Bird, Lerusse, \& Produit}]{goldwurm_integral/ibis_2003}
Goldwurm, A., David, P., Foschini, L., {et~al.} 2003, A\&A, 411,
  \dodoi{10.1051/0004-6361:20031395}

\bibitem[{Greiner {et~al.}(2015)Greiner, Mazzali, Kann, Kr{\"u}hler, Pian,
  Prentice, Olivares~E., Rossi, Klose, Taubenberger, Knust, Afonso, Ashall,
  Bolmer, Delvaux, Diehl, Elliott, Filgas, Fynbo, Graham, Guelbenzu, Kobayashi,
  Leloudas, Savaglio, Schady, Schmidl, Schweyer, Sudilovsky, Tanga, Updike,
  {van Eerten}, \& Varela}]{greiner_very_2015}
Greiner, J., Mazzali, P.~A., Kann, D.~A., {et~al.} 2015, Nat, 523,
  \dodoi{10.1038/nature14579}

\bibitem[{Heussaff {et~al.}(2013)Heussaff, Atteia, \&
  Zolnierowski}]{heussaff_epeak_2013}
Heussaff, V., Atteia, J.-L., \& Zolnierowski, Y. 2013, A\&A, 557,
  \dodoi{10.1051/0004-6361/201321528}

\bibitem[{Holland {et~al.}(2010)Holland, Sbarufatti, Shen, Schady, Cummings,
  Fonseca, Fynbo, Jakobsson, Leitet, Linn{\'e}, Roming, Still, \&
  Zhang}]{holland_grb090417b_2010}
Holland, S.~T., Sbarufatti, B., Shen, R., {et~al.} 2010, ApJ, 717,
  \dodoi{10.1088/0004-637X/717/1/223}

\bibitem[{Hopkins \& Beacom(2006)}]{hopkins_normalization_2006}
Hopkins, A.~M., \& Beacom, J.~F. 2006, ApJ, 651, \dodoi{10.1086/506610}

\bibitem[{Kinugawa {et~al.}(2019)Kinugawa, Harikane, \&
  Asano}]{kinugawa_long_2019}
Kinugawa, T., Harikane, Y., \& Asano, K. 2019, ApJ, 878,
  \dodoi{10.3847/1538-4357/ab2188}

\bibitem[{Kouveliotou {et~al.}(1993)Kouveliotou, Meegan, Fishman, Bhat, Briggs,
  Koshut, Paciesas, \& Pendleton}]{kouveliotou_identification_1993}
Kouveliotou, C., Meegan, C.~A., Fishman, G.~J., {et~al.} 1993, ApJL, 413,
  \dodoi{10.1086/186969}

\bibitem[{Levan(2015)}]{levan_swift_2015}
Levan, A. 2015, JHEAp, 7, \dodoi{10.1016/j.jheap.2015.05.004}

\bibitem[{Levan {et~al.}(2013)Levan, Tanvir, Starling, Wiersema, Page, Perley,
  Schulze, Wynn, Chornock, Hjorth, Cenko, Fruchter, O'Brien, Brown,
  Tunnicliffe, Malesani, Jakobsson, Watson, Berger, Bersier, Cobb, Covino,
  Cucchiara, {de Ugarte Postigo}, Fox, {Gal-Yam}, Goldoni, Gorosabel, Kaper,
  Kr{\"u}hler, Karjalainen, Osborne, Pian, {S{\'a}nchez-Ram{\'i}rez}, Schmidt,
  Skillen, Tagliaferri, Th{\"o}ne, Vaduvescu, Wijers, \&
  Zauderer}]{levan_new_2013}
Levan, A.~J., Tanvir, N.~R., Starling, R. L.~C., {et~al.} 2013, ApJ, 781,
  \dodoi{10.1088/0004-637X/781/1/13}

\bibitem[{Li(2008)}]{li_star_2008}
Li, L.-X. 2008, MNRAS, 388, \dodoi{10.1111/j.1365-2966.2008.13488.x}

\bibitem[{Lien {et~al.}(2016)Lien, Sakamoto, Barthelmy, Baumgartner, Cannizzo,
  Chen, Collins, Cummings, Gehrels, Krimm, Markwardt, Palmer, Stamatikos,
  Troja, \& Ukwatta}]{lien_third_2016}
Lien, A., Sakamoto, T., Barthelmy, S.~D., {et~al.} 2016, ApJ, 829,
  \dodoi{10.3847/0004-637X/829/1/7}

\bibitem[{Mate {et~al.}(2019)Mate, Bouchet, Atteia, Claret, Cordier, Dagoneau,
  Godet, Gros, Schanne, \& Triou}]{mate_simulations_2019}
Mate, S., Bouchet, L., Atteia, J.-L., {et~al.} 2019, Exp Astron, 48,
  \dodoi{10.1007/s10686-019-09643-x}

\bibitem[{Moretti {et~al.}(2009)Moretti, Pagani, Cusumano, Campana, Perri,
  Abbey, Ajello, Beardmore, Burrows, Chincarini, Godet, Guidorzi, Hill, Kennea,
  Nousek, Osborne, \& Tagliaferri}]{moretti_new_2009}
Moretti, A., Pagani, C., Cusumano, G., {et~al.} 2009, A\&A, 493,
  \dodoi{10.1051/0004-6361:200811197}

\bibitem[{Nakauchi {et~al.}(2012)Nakauchi, Suwa, Sakamoto, Kashiyama, \&
  Nakamura}]{nakauchi_long-duration_2012}
Nakauchi, D., Suwa, Y., Sakamoto, T., Kashiyama, K., \& Nakamura, T. 2012, ApJ,
  759, \dodoi{10.1088/0004-637X/759/2/128}

\bibitem[{Nava {et~al.}(2012)Nava, Salvaterra, Ghirlanda, Ghisellini, Campana,
  Covino, Cusumano, D'Avanzo, D'Elia, Fugazza, Melandri, Sbarufatti, Vergani,
  \& Tagliaferri}]{nava_complete_2012}
Nava, L., Salvaterra, R., Ghirlanda, G., {et~al.} 2012, MNRAS, 421,
  \dodoi{10.1111/j.1365-2966.2011.20394.x}

\bibitem[{Racusin \& Burrows(2008)}]{racusin_grb080319b:_2008}
Racusin, J.~L., \& Burrows, D.~N. 2008, AIP Conf. Proc., 1065,
  \dodoi{10.1063/1.3027921}

\bibitem[{Schanne {et~al.}(2015)Schanne, Cordier, Atteia, Godet, Lachaud, \&
  Mercier}]{schanne_eclairs_2015}
Schanne, S., Cordier, B., Atteia, J.-L., {et~al.} 2015, in Proceedings of
  {{Swift}}: 10 {{Years}} of {{Discovery}} \textemdash{} {{PoS}}({{SWIFT}} 10),
  Vol. 233 ({SISSA Medialab}), \dodoi{10.22323/1.233.0107}

\bibitem[{Schanne {et~al.}(2013)Schanne, Provost, Kestener, Gros, Cortial,
  G{\"o}tz, Sizun, Ch{\^a}teau, \& Cordier}]{schanne_scientific_2013}
Schanne, S., Provost, H.~L., Kestener, P., {et~al.} 2013, in 2013 {{IEEE
  Nuclear Science Symposium}} and {{Medical Imaging Conference}},
  \dodoi{10.1109/NSSMIC.2013.6829408}

\bibitem[{Sizun(2011)}]{sizun_synthesis_2011}
Sizun, P. 2011, Synthesis of {{ECLAIRs Geant4}} Simulations, Tech. rep.

\bibitem[{Starling {et~al.}(2011)Starling, Wiersema, Levan, Sakamoto, Bersier,
  Goldoni, Oates, Rowlinson, Campana, Sollerman, Tanvir, Malesani, Fynbo,
  Covino, D'Avanzo, O'Brien, Page, Osborne, Vergani, Barthelmy, Burrows, Cano,
  Curran, De~Pasquale, D'Elia, Evans, Flores, Fruchter, Garnavich, Gehrels,
  Gorosabel, Hjorth, Holland, {van der Horst}, Hurkett, Jakobsson, Kamble,
  Kouveliotou, Kuin, Kaper, Mazzali, Nugent, Pian, Stamatikos, Th{\"o}ne, \&
  Woosley}]{starling_discovery_2011}
Starling, R. L.~C., Wiersema, K., Levan, A.~J., {et~al.} 2011, MNRAS, 411,
  \dodoi{10.1111/j.1365-2966.2010.17879.x}

\bibitem[{Tsvetkova {et~al.}(2017)Tsvetkova, Frederiks, Golenetskii, Lysenko,
  Oleynik, Pal'shin, Svinkin, Ulanov, Cline, Hurley, \&
  Aptekar}]{tsvetkova_konus-wind_2017}
Tsvetkova, A., Frederiks, D., Golenetskii, S., {et~al.} 2017, ApJ, 850,
  \dodoi{10.3847/1538-4357/aa96af}

\bibitem[{{van der Horst} {et~al.}(2014){van der Horst}, Levan, Pooley,
  Wiersema, Kr{\"u}hler, Perley, Starling, Curran, Tanvir, \&
  Wijers}]{van_der_horst_detailed_2014}
{van der Horst}, A.~J., Levan, A.~J., Pooley, G.~G., {et~al.} 2014, MNRAS, 446,
  \dodoi{10.1093/mnras/stu2407}

\bibitem[{Virgili {et~al.}(2013)Virgili, Mundell, Pal'shin, Guidorzi, Margutti,
  Melandri, Harrison, Kobayashi, Chornock, Henden, Updike, Cenko, Tanvir,
  Steele, Cucchiara, Gomboc, Levan, Cano, Mottram, Clay, Bersier, Kopa{\v c},
  Japelj, Filippenko, Li, Svinkin, Golenetskii, Hartmann, Milne, Williams,
  O'Brien, Fox, \& Berger}]{virgili_grb091024a_2013}
Virgili, F.~J., Mundell, C.~G., Pal'shin, V., {et~al.} 2013, ApJ, 778,
  \dodoi{10.1088/0004-637X/778/1/54}

\bibitem[{Waxman {et~al.}(2007)Waxman, M{\'e}sz{\'a}ros, \&
  Campana}]{waxman_grb_2007}
Waxman, E., M{\'e}sz{\'a}ros, P., \& Campana, S. 2007, ApJ, 667,
  \dodoi{10.1086/520715}

\bibitem[{Wei {et~al.}(2016)Wei, Cordier, Antier, Antilogus, Atteia, Bajat,
  Basa, Beckmann, Bernardini, \& Boissier}]{wei_deep_2016}
Wei, J., Cordier, B., Antier, S., {et~al.} 2016.
\newblock \url{https://arxiv.org/abs/1610.06892}

\bibitem[{Zhang {et~al.}(2014)Zhang, Zhang, Murase, Connaughton, \&
  Briggs}]{zhang_how_2014}
Zhang, B.-B., Zhang, B., Murase, K., Connaughton, V., \& Briggs, M.~S. 2014,
  ApJ, 787, \dodoi{10.1088/0004-637X/787/1/66}

\bibitem[{Zhao {et~al.}(2012)Zhao, Cordier, Sizun, Wu, Dong, Schanne, Song, \&
  Liu}]{zhao_influence_2012}
Zhao, D., Cordier, B., Sizun, P., {et~al.} 2012, Exp Astron, 34,
  \dodoi{10.1007/s10686-012-9313-2}

\end{thebibliography}

\appendix

\end{document}